\documentclass[a4paper,11pt]{article}
\usepackage{jheppub}
\usepackage{mathtools}

\newcommand{\beq}{\begin{equation}}
\newcommand{\eeq}{\end{equation}}

\newcommand{\gs}{g_\star}
\newcommand{\gss}{g_{\star s}}
\newcommand{\Trh}{T_\text{RH}}
\newcommand{\arh}{a_\text{RH}}
\newcommand{\rhorh}{\rho_{\rm RH}}
\newcommand{\rhoe}{\rho_{\rm end}}
\newcommand{\aend}{a_\text{end}}
\newcommand{\Tmax}{T_\text{max}}
\newcommand{\rGW}{\rho_\text{GW}}
\newcommand{\rR}{\rho_R}
\newcommand{\rp}{\rho_\phi}
\newcommand{\oGW}{\Omega_\text{GW}}
\newcommand{\DNeff}{\Delta N_\text{eff}}

\newcommand{\mrh}{m_\phi^\text{RH}}
\newcommand{\Hrh}{H_\text{RH}}
\newcommand{\mueff}{\mu_\text{eff}}
\newcommand{\Eom}{E_\omega}

\title{Probing Reheating with Graviton Bremsstrahlung}
\author[a]{Nicolás Bernal,}
\author[b]{Simon Cléry,}
\author[b]{Yann Mambrini}
\author[c]{and Yong Xu\footnote{Corresponding author}}
\affiliation[a]{New York University Abu Dhabi\\
	PO Box 129188, Saadiyat Island, Abu Dhabi, United Arab Emirates}
\affiliation[b]{\it Universit\'e Paris-Saclay, CNRS/IN2P3, IJCLab, 91405 Orsay, France}
\affiliation[c]{\it PRISMA$^+$ Cluster of Excellence and Mainz Institute for Theoretical Physics\\
	Johannes Gutenberg University, 55099 Mainz, Germany}
\emailAdd{nicolas.bernal@nyu.edu}
\emailAdd{clery@ijclab.in2p3.fr}
\emailAdd{mambrini@ijclab.in2p3.fr}
\emailAdd{yonxu@uni-mainz.de}

\abstract{We investigate the stochastic gravitational wave (GW) spectrum resulting from graviton bremsstrahlung during inflationary reheating. We focus on an inflaton $\phi$ oscillating around a generic monomial potential $V(\phi) \propto \phi^n$, considering two different reheating scenarios: $i)$ inflaton decay and $ii)$ inflaton annihilation. 
	We show that in the case of a quadratic potential, the scattering of the inflatons can give rise to larger GW amplitude than the decay channel.
	On the other hand, the GW spectrum exhibits distinct features and redshifts in each scenario, which makes it possible to distinguish them in the event of a discovery. Specifically, in the case of annihilation, the GW frequency can be shifted to values higher than those of decay, whereas the GW amplitude generated by annihilation turns out to be smaller
	than that in the decay case for $n \geq 4$, due to the different scaling of radiation during reheating. We also show that the differences in the GW spectrum become more prominent with increasing $n$. Finally, we highlight the potential of future high-frequency GW detectors to distinguish between the different reheating scenarios.  
}
\begin{document}
	
	\begin{flushright}
		MITP-23-065
	\end{flushright}
	\maketitle
	
	%%%%%%%%%%%%%%%%%%%%%%%%%%%%%%%%%%%%%
	\section{Introduction}
	%%%%%%%%%%%%%%%%%%%%%%%%%%%%%%%%%%%%%
	Cosmic inflation is an elegant paradigm that solves the horizon, flatness, and monopole problems of the (old) standard cosmology~\cite{Lyth:2009zz}. A successful inflationary model must incorporate an efficient reheating mechanism to align with observational data.
	In the conventional scenario, after inflation, the inflaton field descends to the potential minimum, starting to oscillate and transfer energy to light degrees of freedom of the Standard Model (SM), which eventually undergo thermalization, forming the SM bath.
	Energy transfer can proceed through decay or annihilation of the inflaton, facilitated by the introduction of couplings between the inflaton and the daughter fields~\cite{Kofman:1997yn}.
	Due to the inevitable coupling between the metric and the energy-momentum tensor, the emission of graviton degrees of freedom is also anticipated through radiative bremsstrahlung processes~\cite{Weinberg:1965nx, Barker:1969jk}. However, within the framework of general relativity, the graviton production rate is suppressed by the squared Planck mass $M_P^2$, where $M_P= 1/\sqrt{8 \pi\, G_N}\simeq 2.4\times 10^{18}$~GeV. To achieve efficient graviton production, a high-energy scale and consequently a large energy-momentum tensor become imperative. This precise scenario unfolds during the reheating phase.
	
	The physics that takes place during reheating is difficult to test experimentally. One of the very few observables (if not the only one) is gravitational waves (GWs). It is also known that graviton bremsstrahlung during inflationary reheating gives rise to an interesting stochastic GW background (SGWB)\footnote{We notice that fluctuations in the thermal plasma could also give rise to SGWB~\cite{Ghiglieri:2015nfa, Ghiglieri:2020mhm, Ringwald:2020ist, Klose:2022knn, Ringwald:2022xif, Ghiglieri:2022rfp}.}~\cite{Nakayama:2018ptw, Huang:2019lgd, Ghoshal:2022kqp, Barman:2023ymn, Barman:2023rpg, Kanemura:2023pnv}. The spectrum typically peaks at ultrahigh frequencies depending on the detailed dynamics of reheating, such as the shape of the inflaton potential or the inflaton-matter coupling~\cite{Barman:2023ymn, Barman:2023rpg}. 
	It was also recently shown that the GW spectrum can be boosted during bosonic reheating through inflaton decay, once the inflaton oscillates around the minimum of a potential steeper than quadratic, $V(\phi) \propto \phi^n$ with $n > 2$~\cite{Barman:2023rpg}. 
	In this case, the inflaton decay rate has a time dependence proportional to $1/\phi_0^{n-2}$ and $\phi_0^{n-2}$ for bosonic and fermionic reheating, respectively~\cite{Garcia:2020wiy}, where we have assumed that after inflation, $\phi(t) = \phi_0(t)\, \mathcal{P}(t)$, with $\mathcal{P}(t)$ a quasiperiodic function encoding the (an)harmonicity of short-timescale oscillations in the potential and $\phi_0(t)$ is the envelope subject to the effect of redshift and decay.
	Note that $\phi_0(t)$ is a decreasing function of time, leading to the radiation or entropy released in bosonic reheating being smaller compared to that during fermionic reheating. Consequently, the GW amplitude generated in bosonic reheating suffers less dilution effects, and hence tends to feature larger amplitudes.\footnote{Note that the GW amplitude can be written as $\oGW \propto 1/\rR$, where $\rR$ correspond to the energy densities stored in radiation. If $\rR$ is smaller, $\oGW$ can be larger.} 
	
	In the existing literature, the analysis for graviton bremsstrahlung is limited to reheating scenarios considering inflaton decays~\cite{Nakayama:2018ptw, Huang:2019lgd, Ghoshal:2022kqp, Barman:2023ymn, Barman:2023rpg, Kanemura:2023pnv}. In this work, we generalize previous studies by including inflaton annihilation, which is also a viable reheating scenario~\cite{Garcia:2020eof, Garcia:2020wiy, Barman:2021ugy}, and compare their observational signatures in the form of GW emission. In particular, we assume that the inflaton transfers its energy predominantly to the bosonic sector, which could account for the degrees of freedom of the SM Higgs doublet.\footnote{Note that for fermionic reheating, extra vector-like fermionic degree of freedoms beyond the SM of particle physics have to be introduced; in that sense, the bosonic reheating scenario is more minimal.} Although reheating is not possible through inflaton scattering for $n=2$, SM radiation and therefore SM bath temperature feature distinct behaviors depending on the reheating channel for $n>2$.
	More specifically, in the decay scenario, the temperature evolves as $T(a) \propto a^{-\frac{3}{2n+4}}$, while for annihilation $T(a) \propto a^{-\frac{9}{2n+4}}$, where $a$ denotes the cosmic scale factor~\cite{Garcia:2020wiy}.
	We show that the temperature dependence in the redshift is transmitted to the spectrum and amplitude of the GWs generated by the graviton-bremsstrahlung process through the dilution effect, which makes it possible to distinguish between the two reheating channels.
	In particular, we aim to compare the different features of the GW spectrum in inflaton decays and annihilations. This is particularly interesting because it offers the potential to utilize the GW features to pinpoint reheating scenarios with inflaton decays and annihilations. To this end, we first calculate the graviton emission rate from inflaton annihilation.
	Then we compute the bremsstrahlung-induced GW spectrum by solving a set of coupled Boltzmann equations. 
	
	The paper is organized as follows. In Section~\ref{sec:setup}, we study the two reheating setups together with the graviton emission rates in each scenario. In Section~\ref{sec:beq_sol}, the solution for the differential GW spectrum is presented for the inflaton-annihilation scenario. Furthermore, we investigate the GW spectrum, paying particular attention to the difference between inflaton bosonic decay and annihilation. Finally, we summarize our findings in Section~\ref{sec:concl}. 
	
	%%%%%%%%%%%%%%%%%%%%%%%%%%%%%%%%%%%%%
	\section{Reheating and Graviton Emission} \label{sec:setup}
	%%%%%%%%%%%%%%%%%%%%%%%%%%%%%%%%%%%%%
	
	%%%%%%%%%%%%%%%%%%%%%%%%%%%%%%%%%%%%%
	\subsection{Generalities}
	%%%%%%%%%%%%%%%%%%%%%%%%%%%%%%%%%%%%%
	After inflation ends, the inflaton starts to coherently oscillate around the potential minimum, transferring its energy to the SM bath: a process called inflationary reheating~\cite{Kofman:1997yn}. 
	The energy-transfer efficiency is controlled by the type and strength of the inflaton-matter coupling. In this paper, we focus on a bosonic reheating scenario, considering that the inflaton $\phi$ decays or annihilates into a real scalar $\varphi$, which could account for the degrees of freedom of the SM Higgs doublet.
	
	The relevant interactions are described by the Lagrangian density
	\begin{equation}
	\mathcal{L} \supset \mu\, \phi\, \varphi^2  + \sigma\, \phi^2\, \varphi^2,
	\end{equation}
	where $\mu$ and $\sigma$ are couplings with mass dimensions 1 and 0, respectively.
	Expanding the space-time metric $g_{\mu \nu}$ around the flat Minkowski background $g_{\mu \nu} \simeq \eta_{\mu \nu} + (2/M_P)\, h_{\mu \nu}$, one can write the effective coupling~\cite{Choi:1994ax}
	\begin{equation}
	\sqrt{-g}\,  \mathcal{L} \supset -\frac{1}{M_P}\, h_{\mu \nu}\, T^{\mu \nu},
	\end{equation}
	where $g$ denotes the determinant of $g_{\mu \nu}$ and $T^{\mu \nu}$ corresponds to the energy-momentum tensor.
	
	During reheating, we assume that the inflaton oscillates around a generic monomial potential of the form~\cite{Turner:1983he, Garcia:2020wiy, Bernal:2022wck}
	\begin{equation} \label{eq:potential}
	V(\phi) = \lambda\, M_P^4 \left(\frac{\phi}{M_P}\right)^n,
	\end{equation}
	where $\lambda$ is a dimensionless parameter.
	Such potential could originate, for example, from Starobinsky inflation~\cite{Starobinsky:1980te}, polynomial inflation~\cite{Drees:2021wgd, Bernal:2021qrl, Drees:2022aea} or the $\alpha$-attractor inflationary $T$- or $E$-models~\cite{Kallosh:2013hoa, Kallosh:2013yoa}.
	The effective mass $m_\phi$ 
	of the inflaton can be defined as the second derivative of its potential and is~\cite{Garcia:2020wiy, Bernal:2022wck}
	\begin{equation} \label{eq:m2}
	m_\phi(a)^2 \simeq n\, (n-1)\, \lambda^\frac{2}{n}\, M_P^\frac{2\, (4 - n)}{n} \rp(a)^{\frac{n-2}{n}},
	\end{equation}
	which is a time-dependent quantity for $n \neq 2$ given as a function of the inflaton energy density
	\begin{equation}
	\rho_\phi = \frac12\, \dot \phi^2 + V(\phi)\,,
	\end{equation}
	where dots $(\dot {\phantom .})$ correspond to derivatives with respect to time.
	
	The evolution of the energy densities for the inflaton $\rp$ and the SM radiation $\rR$ can be tracked by the following Boltzmann equations~\cite{Turner:1983he, Bernal:2022wck, Garcia:2020wiy}
	\begin{align} 
	\frac{d\rp}{dt} +  3H\, (1+w_\phi) \rp &= - (1+w_\phi)\, \gamma\,, \label{eq:drPdt}\\
	\frac{d\rR}{dt} + 4\, H\, \rR &= +  (1+w_\phi)\, \gamma \label{eq:drRdt}\,,
	\end{align}
	where $w_\phi = (n-2)/(n+2)$ denotes the equation of state~\cite{Bernal:2022wck, Garcia:2020wiy} and $\gamma \equiv \Gamma\, \rho_\phi$ is the interaction rate density for the production of SM states from inflatons. Additionally, $H$ corresponds to the Hubble expansion rate
	\begin{equation}
	H^2 = \frac{\rR + \rp}{3\, M_P^2}\,,
	\end{equation}
	with
	\begin{equation}
	\rR(T) = \frac{\pi^2}{30}\, \gs(T)\, T^4,
	\end{equation}
	and $\gs(T)$ being the number of relativistic degrees of freedom contributing to the SM energy density.\footnote{Note that during our study, we assumed instantaneous thermalization, for details of the effects of non-instantaneous thermalization, see Refs.~\cite{Mukaida:2015ria, Garcia:2018wtq, Chowdhury:2023jft}.}
	The end of the reheating period is defined as the moment at which the SM radiation starts to dominate the total energy density of the universe and corresponds to a reheating temperature $\Trh$ (or equivalently, a scale factor $\arh \equiv a(\Trh)$) given by $\rR(\Trh) = \rp(\Trh) =3\, M_P^2\, \Hrh^2$, where $\Hrh \equiv H(\Trh)$.
	
	During reheating, that is, when $\Gamma \ll H$, Eq.~\eqref{eq:drPdt} admits an approximate analytical solution given by
	\begin{equation} \label{eq:rp}
	\rp(a) \simeq \rp (\arh) \left(\frac{\arh}{a}\right)^\frac{6\, n}{n+2},
	\end{equation}
	and therefore, the effective inflaton mass can be rewritten as
	\begin{equation}
	m_\phi(a) = \mrh \left(\frac{\arh}{a}\right)^\frac{3\, (n - 2)}{n + 2},
	\end{equation}
	with 
	\begin{equation} \label{eq:Mrh}
	\mrh \equiv m_\phi(\arh) \simeq \sqrt{n\, (n-1)}\, \lambda^\frac{1}{n}\, M_P \left(\sqrt{3}\, \frac{\Hrh}{M_P}\right)^{\frac{n-2}{n}}.
	\end{equation}
	Depending on the form of $\gamma$, the solution for $\rR(a)$ and therefore the temperature $T(a)$ during reheating exhibits distinct characteristics, which would consequently give rise to different dilution effects for the generated GW spectrum. In the following sections, we will present the scenarios with inflaton bosonic decay and annihilation.
	
	However, before proceeding, it is important to comment on several points. We have assumed that the effects of non-perturbative preheating~\cite{Amin:2014eta, Garcia:2021iag}, inflaton fragmentation~\cite{Lozanov:2016hid, Garcia:2023eol, Garcia:2023dyf}, and gravitational reheating~\cite{Bernal:2020qyu, Clery:2021bwz, Haque:2022kez, Barman:2023opy, Haque:2023zhb} are subdominant compared to perturbative decay. Note that even if non-perturbative phenomena during reheating could lead to the equation of state approaching $1/3$ during reheating, perturbative decay is still necessary to fully deplete the inflaton energy~\cite{Dufaux:2006ee, Maity:2018qhi}. 
	Furthermore, for the bosonic decay scenario, the inefficiency of preheating has been highlighted as the result of self-interaction of the generated daughter field~\cite{Dufaux:2006ee}. In particular, the trilinear coupling $\mu\, \phi\, \varphi^2$ gives rise to a tachyonic squared mass $m^2_{\varphi} \sim \mu\, \phi$ for the field $\varphi$ once $\phi$ crosses zero and becomes negative. However, the self-interaction of the $\varphi$ fields $\lambda_\varphi\, \varphi^4$ gives rise to a positive squared mass term $m^2_{\varphi} \sim \lambda_{\varphi} \left\langle \varphi^2 \right \rangle $ where $\left\langle \varphi^2 \right \rangle $ corresponds to the variance of $\varphi$. This backreaction counteracts tachyonic effects and quickly terminates preheating, making it inefficient~\cite{Dufaux:2006ee}. 
	Additionally, when the inflaton features self-interactions or, equivalently, when the inflaton potential is steeper than quadratic, the effects of inflaton fragmentation are relevant. Indeed, in the case of decays into fermions, fragmentation is particularly important, making perturbative decays subdominant~\cite{Garcia:2023dyf}.
	However, in the case of decay or scattering of the inflaton to scalars, the effect is milder.
	
	Lastly, to be dominant, the gravitational reheating mechanism requires $w_\phi \gtrsim 0.65$, corresponding to $n > 9$~\cite{Clery:2021bwz, Haque:2022kez, Co:2022bgh, Barman:2022qgt}. This bound can be relaxed to $n > 4$ if one introduces a non-minimal coupling between gravity and a pair of inflatons~\cite{Clery:2022wib}, or even to $n \geq 2$ if gravity couples non-minimally to a single inflaton~\cite{Barman:2023opy}.
	In the current work, we focus on scenarios with $2 \leq n \leq 6$ minimally coupled to gravity with high-temperature reheating, where the effects of inflaton fragmentation and gravitational reheating are expected to be subdominant.
	
	%%%%%%%%%%%%%%%%%%%%%%%%%%%%%%%%%%%%%%%%%%%%%%%%%%%%%%
	\subsection{Graviton emission from inflaton decay} \label{sec:inflaton_dec}
	%%%%%%%%%%%%%%%%%%%%%%%%%%%%%%%%%%%%%%%%%%%%%%%%%%%%%%
	In this section, we briefly review the case of inflaton bosonic decays~\cite{Barman:2023rpg, Barman:2023ymn}.
	The rate of inflaton decay into a pair of real scalars $\varphi$ with mass $m$ is given by
	\begin{equation} \label{eq:Gamma_decay}
	\Gamma^{1\to2} \simeq \frac{m_\phi}{8\pi} \left(\frac{\mueff}{m_\phi}\right)^2,
	\end{equation}
	where
	\beq
	\mueff^2(n) \simeq (n+2) (n-1)\, \mu^2\, \frac{\varpi}{m_\phi} \sum_{j=1}^{\infty} j\, |\mathcal{P}_j|^2\left\langle\left( 1-\frac{4y^2}{j^2}\mathcal{P}\right)^{1/2} \right \rangle
	\eeq
	corresponds to the effective coupling after averaging over inflaton oscillations~\cite{Garcia:2020wiy}, and $y \equiv m/\varpi$. We introduced the Fourier decomposition of the oscillating field
	\begin{equation}
	\phi(t) = \phi_0(t)\, \mathcal{P}(t) = \phi_0(t) \sum_{j=1}^{\infty} \mathcal{P}_j(t)\, e^{-i\, j\, \varpi\, t}.
	\end{equation}
	The $j^\text{th}$ oscillating mode is associated with an energy $E_j =j\, \varpi$ where $\varpi$ is the frequency of the oscillating background field. It is a time-dependent quantity related to the effective inflaton mass and amplitude via 
	\beq
	\varpi = \sqrt{\frac{n\, \pi}{2 (n-1)}}\, \frac{\Gamma(\frac{1}{2} + \frac{1}{n})}{\Gamma(\frac{1}{n})}\, m_\phi = \alpha_n\, \sqrt{\lambda}\, M_P \left(\frac{\phi_0}{M_P}\right)^\frac{n-2}{2}.
	\eeq
	We note that the energy in each mode is time dependent for $n>2$ and that each mode contributes to the production rate, as can be seen from the sum in the expression above. In what follows, we neglect the contribution of higher-order modes $j>1$ and look at the leading contribution $j=1$.
	
\begin{figure}[t!]
		\def\sepf{0.59}
		\centering
		\includegraphics[scale=\sepf]{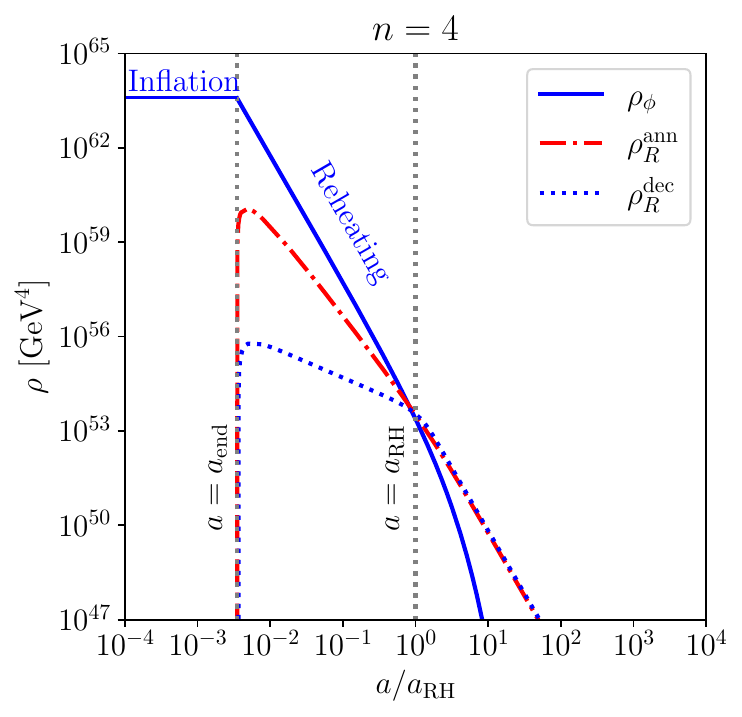}
		\includegraphics[scale=\sepf]{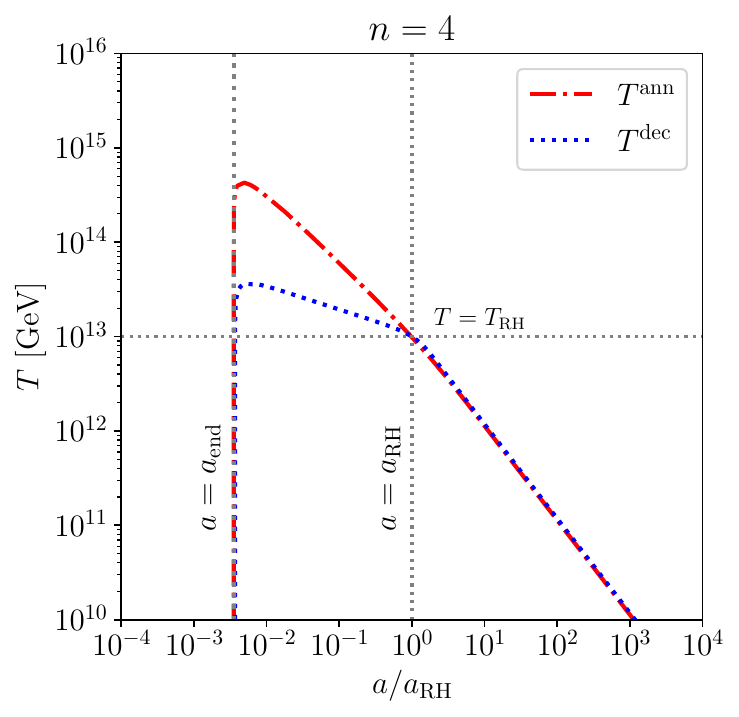}\\
		\includegraphics[scale=\sepf]{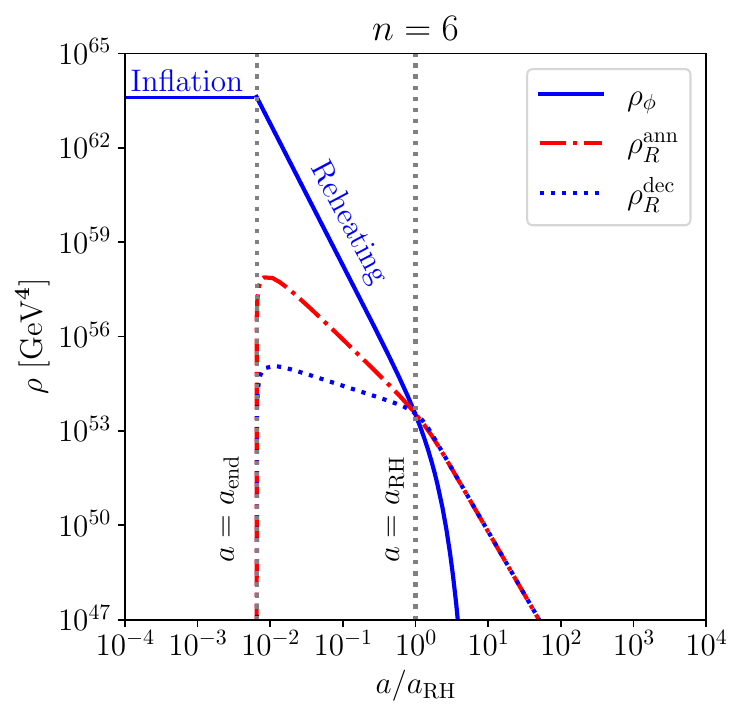}
		\includegraphics[scale=\sepf]{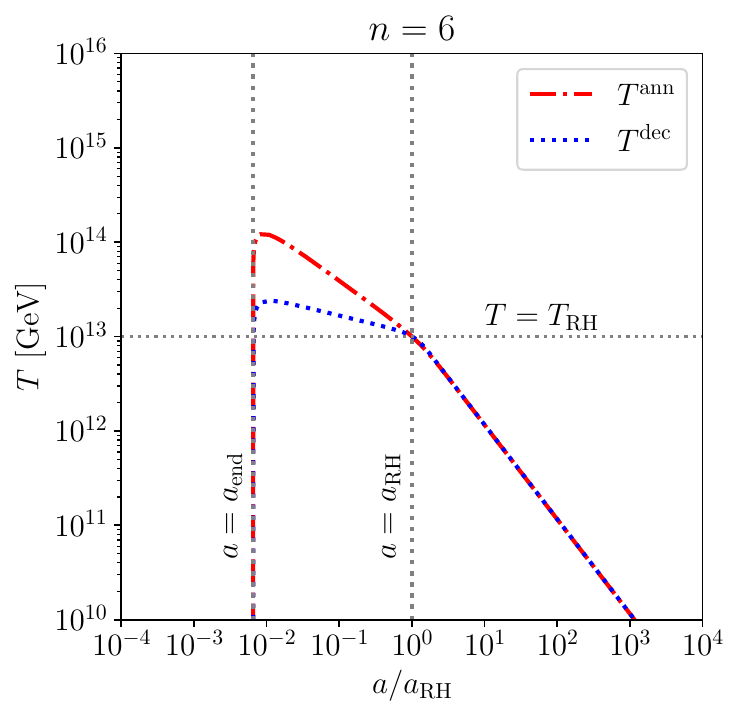}\\
		\caption{Evolution for energy densities (left) and temperature (right), for $\Trh = 10^{13}$~GeV. In the upper panels, we have chosen $n = 4$ and $\Tmax = 3.6 \times 10^{13}$~GeV for decays ($\Tmax = 4.3 \times 10^{14}$~GeV for annihilations), while in the lower panels $n = 6$ and $\Tmax = 2.4 \times 10^{13}$~GeV for decays ($\Tmax = 1.2 \times 10^{14}$~GeV for annihilations).
		}
		\label{fig:rho_T}
	\end{figure}

	In the case of decays, the rate density $\gamma = \rp\, \Gamma^{1\to2}$, and therefore Eq.~\eqref{eq:drRdt} admits the analytical solution~\cite{Xu:2023lxw}
	\begin{equation} \label{eq:rR_dec}
	\rR(a) \simeq \frac{n }{1+2n} \rp(\aend) \left(\frac{\aend}{a}\right)^{\frac{6}{n+2}}\left[1-\left(\frac{\aend}{a}\right)^{\frac{2+4n}{2+n}}\right]  \left(\frac{\aend}{\arh}\right)^{\frac{6(n-1)}{2+n}},
	\end{equation}
	where $\aend$ and $\arh$  denote the scale factor at the end of inflation and of reheating, respectively, which implies that during reheating
	\begin{equation}
	T(a) = \Trh \left(\frac{\arh}{a}\right)^\alpha,
	\end{equation}
	with
	\begin{equation}
	\alpha \equiv \frac{3}{2n+4}\,.
	\end{equation}
	It is important to note that away from the instantaneous decay approximation of the inflaton, the SM bath could reach temperatures higher than $\Trh$, up to $\Tmax$~\cite{Giudice:2000ex}. This can be observed in Fig.~\ref{fig:rho_T}, where we show the evolution of the energy densities of the inflaton and SM radiation (left) and the temperature of the SM bath (right) as a function of the scale factor for the cases of inflaton annihilations and decays. Reheating occurs in the range $\aend < a < \arh$. Earlier times correspond to the cosmic inflationary epoch, while later times to the SM radiation-dominated isentropic era, where $T(a) \propto a^{-1}$. We note that to solve Eqs.~\eqref{eq:drPdt} and~\eqref{eq:drRdt}, one has to fix the interaction rate density $\gamma$ of the inflaton (and, therefore, its mass and couplings to the daughter particles), its initial energy density $\rp(\aend)$ and $n$. We recall that at the end of inflation $\rR(\aend) = 0$ is expected. However, it is more convenient to choose, instead of $\gamma$ and $\rp(\aend)$ as free parameters, $\Trh$ and $\Tmax$. The latter set of parameters is used in the following analysis.
	In the upper panels of Fig.~\ref{fig:rho_T}, we have chosen $n = 4$, $\Trh = 10^{13}$~GeV with $\Tmax = 3.6 \times 10^{13}$~GeV, while in the lower panels $n = 6$, $\Trh = 10^{13}$~GeV with $\Tmax = 2.4 \times 10^{13}$~GeV. We consider $\Trh$ as a free parameter and then compute $\Tmax$ in the $\alpha$-attractor $T$-model considering the central values of the amplitude of the primordial spectrum of curvature perturbations and its associated spectral tilt: $n_s=0.9659$ and $A_s=2.1\times 10^{-9}$~\cite{Planck:2018vyg}.
	%%%%%%%%%%%%%%%%%%%%%%%%%%%%%%%%%%%%%%%%%%
	%%%%%%%%%%%%%%%%%%%%%%%%%%%%%%%%%%%%%%%%%%
	
	%%%%%%%%%%%%%%%%%%%%%%%%%%%%%%%%%%%%%%%%%%
	\begin{figure}[t!]
		\def\sepf{0.4}
		\centering
		\includegraphics[scale=\sepf]{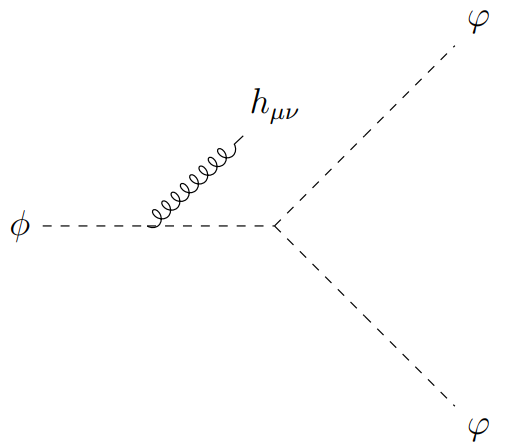}
		\includegraphics[scale=\sepf]{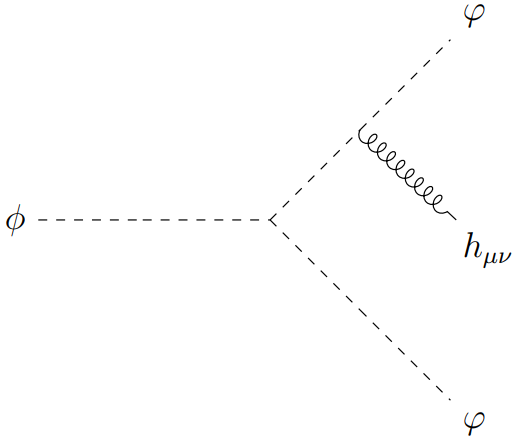}
		\includegraphics[scale=\sepf]{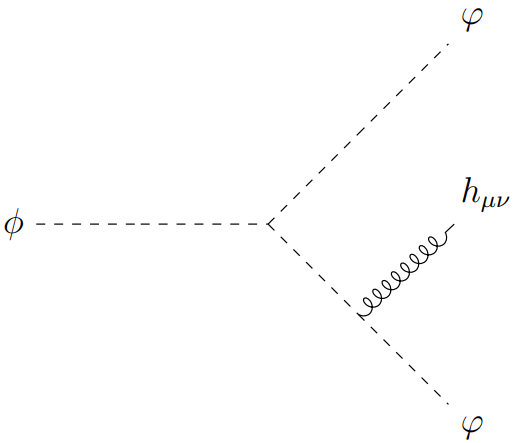}
		\includegraphics[scale=\sepf]{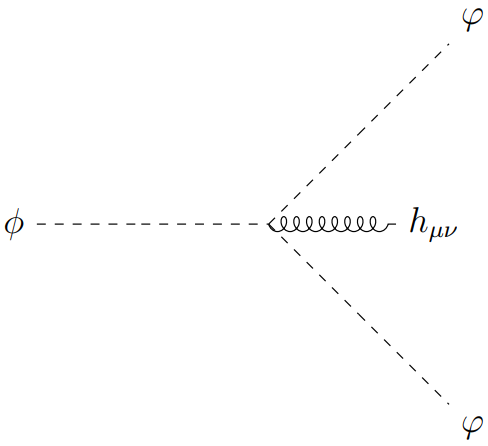}
		\caption{Inflaton decays into a pair of $\varphi$ and a graviton $h_{\mu\nu}$.} 
		\label{fig:3body}
	\end{figure}
	%%%%%%%%%%%%%%%%%%%%%%%%%%%%%%%%%%%%%%%%%%
	In addition to the 2-body decay, there is an inexorably associated graviton emission by bremsstrahlung, as shown in Fig.~\ref{fig:3body}. It is interesting to note that from the four possible diagrams, two of the matrix elements (the first and the last) vanish because the inflaton decays with vanishing three momenta and the graviton polarization tensor is traceless~\cite{Barman:2023ymn}.
	The differential decay width for the inflaton condensate with respect to the graviton energy $\Eom$ for this 3-body decay process is given by~\cite{Garcia:2020wiy}
	\beq
	\frac{d\Gamma^{1\to3}}{d\Eom} =\frac{m_\phi}{\rho_\phi(1+w_\phi)} \sum\limits_{j=1}^{\infty} \int \left|\mathcal{M}^{1\to3}_j\right|^2 \frac{d\rm PS^{(3)}}{d\Eom}\,,
	\label{Eq:1to3diffwidth}
	\eeq
	where $d\rm PS^{(3)}$ is the 3-body Lorentz-invariant phase space, leading to\footnote{Note that this is consistent with the result presented in Ref.~\cite{Barman:2023ymn} where it was assumed that the final state contains different particles and that there are two decay channels. The prefactor in Ref.~\cite{Barman:2023ymn} is $1/32$, which recovers to $1/16$ after multiplying  $2^2/2$.}
	\beq \label{eq:dGdE}
	\frac{d\Gamma^{1\to3}}{d\Eom} \simeq \frac{1}{16\pi^3} \left(\frac{\mu_{\rm eff}} {M_P}\right)^2 \left[\frac{(1 - 2 x)\, (1 - 2 x + 2 y^2)}{4\, x\, \zeta^{-1}} + \frac{y^2\, (y^2 + 2 x - 1)}{x}\ln\left(\frac{1 + \zeta}{1 - \zeta}\right)\right],
	\eeq
	with $x \equiv \Eom/\varpi$, and 
	\begin{equation}
	\zeta \equiv \sqrt{1 - \frac{4\, y^2}{1-2\, x}}\,.
	\end{equation}
	In the above expression, we have considered only the contribution of the first mode $j=1$ in $\mu_{\rm eff}$.
	The graviton energy spans the range
	\begin{equation}\label{eq:EoM_decay}
	0 < \Eom \leq m_\phi \left(\frac12 - 2\, y^2\right).
	\end{equation}
	We see that when $\Eom \to m_\phi \left(\frac12 - 2\, y^2\right)$, the decay becomes kinematically forbidden,\footnote{It was shown in Ref.~\cite{Mambrini:2022uol} that graviton bremsstrahlung can also be the source of dark matter production.} and therefore $d\Gamma^{1\to3}/d\Eom \to 0$. Note that here $\Eom $ corresponds to the graviton energy at emission, which can be at most half of the inflaton mass in the limit $y \to 0$.
	
	%%%%%%%%%%%%%%%%%%%%%%%%%%%%%%%%%%%%%%%%%%%%%%%%%%%%%%
	\subsection{Graviton emission from inflaton annihilation} \label{sec:inflaton_ann}
	%%%%%%%%%%%%%%%%%%%%%%%%%%%%%%%%%%%%%%%%%%%%%%%%%%%%%%
	Instead of decay, the transfer of energy to the SM bath during reheating could be dominated by inflaton annihilation, see e.g. Refs.~\cite{Hooper:2018buz, Garcia:2021gsy} for motivated scenarios.
	The $2\to2$ annihilation cross section of inflatons in a pair of real scalars $\varphi$, in the non-relativistic limit is given by
	\beq
	\Gamma^{2\to2} \simeq \frac{\sigma^2_{\rm eff}\, \rho_\phi}{8\pi\, m_\phi^3}\,,
	\eeq
	where 
	\beq
	\sigma_{\text{eff}}^2  \simeq \sigma^2\, n\, (n+2) (n-1)^2\,\frac{\varpi}{m_\phi}\, \sum_{j=1}^{\infty} j\, |\mathcal{P}_j|^2 \left\langle\left(1 - \frac{4\, y^2}{j^2}\, \mathcal{P}\right)^{1/2} \right\rangle
	\eeq
	denotes the effective coupling after averaging inflaton oscillations~\cite{Garcia:2020wiy}. We have introduced the Fourier decomposition of the oscillating background field square
	\begin{equation}
	\phi(t)^2 = \phi_0^2(t)\, \mathcal{P}^{(2)}(t) = \phi_0^2(t)\, \sum\limits_{j=1}^{\infty} \mathcal{P}^{(2)}_j(t)\, e^{-i\, j\, \varpi\, t}.
	\end{equation}
	The corresponding interaction rate density is $\gamma \equiv \Gamma^{2\to 2}\, \rho_\phi$, and therefore Eq.~\eqref{eq:drRdt} admits the analytical solution~\cite{Xu:2023lxw}
	\begin{equation}\label{eq:rR_ann}
	\rR(a) \simeq \frac{n}{2n-5}\, \rp(\aend) \left(\frac{\aend}{a}\right)^{\frac{18}{n+2}} \left[1 - \left(\frac{\aend}{a}\right)^{\frac{2(2n-5)}{2+n}}\right] \left(\frac{\aend}{\arh}\right)^{\frac{6(n-3)}{n+2}},
	\end{equation}
	which in turn implies that during reheating
	\begin{equation}
	\alpha \equiv \frac{9}{2\, n+4}\,,
	\end{equation}
	for $n \geq 5/2$.
	Interestingly, for $n = 2$ (and, in general, for $n < 5/2$), the radiation energy density dilutes faster than nonrelativistic matter (i.e. the inflaton), which implies that it cannot overcome the inflaton energy density, and, therefore, the universe cannot become radiation dominated. In the case of annihilation, Fig.~\ref{fig:rho_T} also shows the evolution of the energy densities (left) and the temperature of the SM (right) as a function of the scale factor, for $\Trh = 10^{13}$~GeV, and $\Tmax = 4.3 \times 10^{14}$~GeV for $n=4$ (top), or $\Tmax = 1.2  \times 10^{14}$~GeV for $n=6$ (bottom).
	It is clear that in the annihilation case, the evolution of radiation or temperature is steeper than that in the decay case, which has important implications for the GW spectrum, as will be shown in the next section.
	
	There are some self-consistency conditions that must be met.
	First, one has to ensure that the inflaton mass satisfies $m_\phi(a) < M_P$~\cite{Barman:2023rpg}, which leads to upper bounds on the reheating energy scale: $\mrh < M_P\, (\Tmax/\Trh)^{2(2-n)/3}$ for annihilations and $\mrh < M_P\, (\Tmax/\Trh)^{2(2-n)}$ for decays. Alternatively, if one treats $\mrh$ as a free parameter, the condition $m_\phi(a) < M_P$ also gives upper bounds on the temperature ratio, which reads $(\Tmax/\Trh)< (\mrh/M_P)^{3/(4-2n)}$ for annihilations and $(\Tmax/\Trh) < (\mrh/M_P)^{1/(4-2n)}$ for decays. Furthermore, we can take advantage of the recent constraint on the inflationary tensor-to-scalar ratio $r < 0.035$, obtained from BICEP/Keck 2018~\cite{BICEP:2021xfz}, to derive an upper bound on the inflationary scale, denoted by $H_I < 2.0 \times 10^{-5}~M_P$. This upper bound on the inflationary scale, in turn, allows us to establish upper bounds on the reheating temperature. Consequently, for the scenario of annihilation, the upper bounds on the reheating temperature are given by $\Trh < 3.3 \times 10^{15}~\text{GeV}\, (\Tmax/\Trh)^{-4/3}$ with $n = 4$ and $\Trh < 3.1 \times 10^{15}$~GeV $(\Tmax/\Trh)^{-2}$ with $n = 6$. On the other hand, for the inflaton decay scenario, the bounds on the reheating temperature are expressed as $\Trh < 2.8 \times 10^{15}~\text{GeV}\, (\Tmax/\Trh)^{-4}$ for $n = 4$ and $\Trh < 2.5 \times 10^{15}~\text{GeV}\, (\Tmax/\Trh)^{-6}$ for $n = 6$. We have checked that the values of $\Tmax$ and $\Trh$ shown in Fig.~\ref{fig:rho_T} satisfy these constraints.
	
	%%%%%%%%%%%%%%%%%%%%%%%%%%%%%%%%%%%%%%%%
	\begin{figure}[t]
		\def\sepf{0.8}
		\centering
		\includegraphics[scale=\sepf]{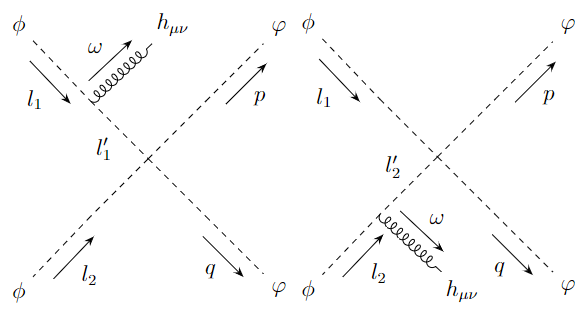}
		\includegraphics[scale=\sepf]{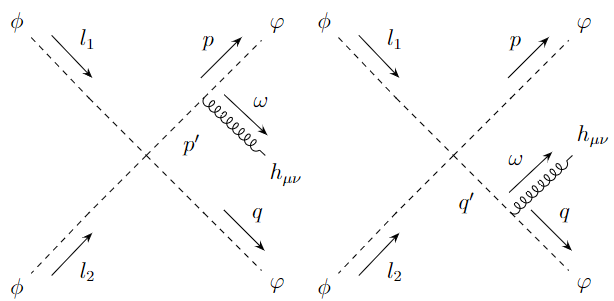}
		\includegraphics[scale=\sepf]{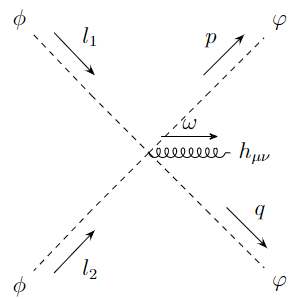}
		\caption{Inflaton annihilation into a pair of $\varphi$ and a graviton.}
		\label{fig:anni}
	\end{figure}
	%%%%%%%%%%%%%%%%%%%%%%%%%%%%%%%%%%%%%%%%
	Similarly to the case of decay, there is a $2 \to 3$ inflaton annihilation process containing the emission of a graviton, as shown in Fig.~\ref{fig:anni}. Using the Feynman rules and the formalism as presented in Ref.~\cite{Barman:2023ymn}, we obtain the following amplitudes
	\begin{align}
	i\mathcal{M}^1_j &= -\, \frac{i\, \sigma\phi_0^2}{M_P}\, \frac{l_{1\,\mu}\, l_{1\,\nu}\, \epsilon^{\star\mu\nu}}{l_1 \cdot \omega}\,\mathcal{P}_j^{(2)}, \label{ana_M1} \\
	i\mathcal{M}^2_j &= -\frac{i\, \sigma\phi_0^2}{M_P}\, \frac{l_{2\,\mu}\, l_{2\,\nu}\epsilon^{\star\mu\nu}}{l_2\cdot \omega}\,\mathcal{P}_j^{(2)}, \label{ana_M2} \\
	i\mathcal{M}^3_j &= \frac{i\, \sigma\phi_0^2}{M_P}\, \frac{p_\mu\, p_\nu\, \epsilon^{\star\mu\nu}}{p\cdot \omega}\,\mathcal{P}_j^{(2)}, \label{ana_M3}\\
	i\mathcal{M}^4_j &= \frac{i\, \sigma\phi_0^2}{M_P}\, \frac{q_\mu\, q_\nu\, \epsilon^{\star\mu\nu}}{q\cdot \omega}\,\mathcal{P}_j^{(2)},\label{ana_M4}\\
	i\mathcal{M}_5 & \propto \eta_{\mu \nu} \,\epsilon^{\star\mu\nu}, \label{ana_M5}
	\end{align}
	where
	\begin{align}
	p \cdot \omega &= 2 m_\phi \left(\Eom + E_p -  m_\phi \right),\\
	q \cdot \omega &=(l_1\, + l_2- \omega - p ) \cdot \omega = 2 m_\phi\, \Eom -p  \cdot \omega\,,
	\end{align}
	and $\epsilon^{\mu\nu}$ are the polarisation tensors of the graviton.
	Note that $\mathcal{M}_5$ vanishes due to the traceless condition of the massless graviton, while $\mathcal{M}_1$ and $\mathcal{M}_2$ vanish since we assume that the inflaton condensate annihilates with a vanishing three-momentum~\cite{Barman:2023ymn}.
	
	The $2\to3$ differential interaction rate reads\footnote{Note that there we have $2\, m_\phi$ since the initial energy for annihilation is twice larger compared to that from the decay, which is defined in Eq.~\eqref{Eq:1to3diffwidth}.}
	\beq
	\frac{d\Gamma^{2\to3}}{d\Eom} = \frac{2\,m_\phi}{\rho_\phi (1+w_\phi)}\sum\limits_{j=1}^{\infty}\int|\mathcal{M}^{2\to3}_j|^2\frac{d\rm PS^{(3)}}{d\Eom}\,,
	\eeq
	leading to
	\beq \label{eq:2to3diffrate}
	\frac{d\Gamma^{2\to3}}{d\Eom} \simeq \frac{\sigma^2_{\rm eff}}{16\pi^3} \frac{\rho_\phi}{m_\phi^2 M_P^2} \left[\frac{(1-x)\, (2-2x +y^2)}{x\, \beta^{-1}} + \frac{ y^2(y^2 + 4x - 4)}{2x} \log\left(\frac{1 + \beta}{1 - \beta}\right)\right],
	\eeq
	where we have considered only the contribution of the first mode in $\sigma_{\rm eff}$, with
	\begin{equation}
	\beta \equiv \sqrt{1 - \frac{y^2}{1-x}}\,,
	\end{equation}
	and the graviton energy in the range
	\begin{equation}
	0 < \Eom \leq m_\phi \left(1 -  y^2\right).
	\end{equation}
	As expected, in the limit $E_\omega \to m_\phi(1-y^2)$, there is no graviton emission as $2 \to 3$ annihilations become kinematically forbidden. Note that for annihilation, the graviton energy $E_\omega$ can be as large as the inflaton mass in the limit $y \to 0$, which is twice larger than that from the decay (cf. Eq.~\eqref{eq:EoM_decay}). We will see in the next section that such distinctions have important implications for the range of frequency for the emitted GWs in the two processes.
	
	%%%%%%%%%%%%%%%%%%%%%%%%%%%%%%%%%%%%%%%%%%%%%%%%%%%%%%%%%%%%%%%
	\section{Gravitational Wave Spectrum} \label{sec:beq_sol}
	%%%%%%%%%%%%%%%%%%%%%%%%%%%%%%%%%%%%%%%%%%%%%%%%%%%%%%%%%%%%%%%
	
	%%%%%%%%%%%%%%%%%%%%%%%%%%%%%%%%%%%%%%%%%%%%%%%%%%%%%%%%%%%%%%%
	\subsection{Overview}
	%%%%%%%%%%%%%%%%%%%%%%%%%%%%%%%%%%%%%%%%%%%%%%%%%%%%%%%%%%%%%%%
	Before analyzing in more detail the spectrum and constraints of GWs generated by brems\-strahlung of gravitons, in this section we summarize the procedure in the case of inflaton decay and scattering, in the simplest case of a quadratic potential $V(\phi)=\frac{1}{2} m_\phi^2\, \phi^2$. We refer the reader to Appendix~\ref{Sec:appgenericV} for the complete treatment, which takes into account the sum over all modes in the case of generic potentials. Although we know that complete reheating is not possible in this scenario for pure scattering~\cite{Garcia:2020eof, Garcia:2020wiy}, this example will help us better understand the context and expected results. Moreover, it is always possible to imagine mixed scenarios, where graviton emission by scattering is independent of the process responsible for reheating, such as gravitational reheating.
	
	The differential Boltzmann equation for the energy density $\rGW$ stored in the form of graviton radiation generated by inflaton decay is
	\beq
	\frac{d}{dt} \left(\frac{d \rGW}{d\Eom}\right) + 4\, H\, \frac{d\rGW}{d\Eom} = 
	\frac{\rho_\phi}{m_\phi}\, \frac{d\Gamma^{1 \rightarrow 3}}{d\Eom} \times \Eom\,,
	\label{Eq:friedmann}
	\eeq
	where $d\Gamma^{1\to 3}/d\Eom$ is given by Eq.~\eqref{eq:dGdE}.
	This equation can be rewritten as
	\beq
	\frac{d}{da}\left(a^4\, \frac{d \rGW}{d\Eom}\right) = \frac{a^3}{H}\, \frac{\rho_\phi}{m_\phi}\, \frac{d \Gamma^{1 \rightarrow 3}}{d\Eom} \times \Eom\,,
	\label{eq:drhoGWdE}
	\eeq
	which, in the limit $m$, $\Eom \ll m_\phi$, admits the approximate analytical solution
	\beq
	\left.\frac{d\rGW}{d\Eom}\right|_{\arh} \simeq \frac{\sqrt{3}}{160 \pi^3}\, \frac{\mu^2\sqrt{\rhorh}}{M_P}\,,
	\eeq
	where we used $\rho_\phi(a) = \rhorh \left(\frac{\arh}{a}\right)^3$ in the case of a quadratic potential.
	To obtain the gravitational energy density today $\oGW^0$, at energy $E_\omega$, we need to apply the redshift $(\arh/a_0)^4$ for the energy density and $\arh/a_0$ for the graviton energy. 
	We then have 
	\beq
	\oGW^0 = \frac{1}{\rho_c^0}\, 
	\left.\frac{d \rGW}{d \ln \Eom}\right|_{a_0} = \frac{\sqrt{3}}{160 \pi^3}\, \frac{\mu^2\, \rho_R^0}{M_P\, \rho_c^0\, \sqrt{\rhorh}} \times \Eom\,,
	\label{Eq:omegadecay}
	\eeq
	where $\Omega_R^0=\frac{\rho_R^0}{\rho_c^0}$ is the relative radiation density today, $\rho_c^0 \simeq 1.05 \times 10^{-5}~h^2$~GeV/cm$^3$ the critical density at present, and we have used $\arh/a_0 = (\rho_R^0/\rhorh)^{1/4}$. Alternatively, expressing Eq.~\eqref{Eq:omegadecay} as function of the frequency defined by $\Eom(a) = 2 \pi\, f\, \frac{a_0}{a}$, we obtain 
	\beq
	\oGW^{1\rightarrow 3} h^2 \simeq  10^{-20} \left(\frac{10^{10}~\text{GeV}}{\Trh}\right) \left(\frac{\mu}{10^{10}~\text{GeV}}\right)^2 \left(\frac{f}{10^{8}~\rm{Hz}}\right),
	\label{Eq:spectrumdecay}
	\eeq
	with $h\simeq 0.67$ the present dimensionless Hubble parameter.
	Now, if one considers that the reheating is {\it also} given by the same decaying channel, we can express the GW spectrum completely as a function of $\Trh$ (or $\mu$) considering $\Gamma^{1\to 2} = \frac{\mu^2}{8 \pi\, m_\phi}$ from Eq.~\eqref{eq:Gamma_decay}, and $\rhorh \simeq \Gamma_\phi^2\, M_P^2$.
	In this case, $\mu = 10^{10}$~GeV gives $\Trh \sim 10^{12}$~GeV with $m_\phi \simeq 10^{13}~\text{GeV}$. We note that Eq.~\eqref{Eq:spectrumdecay} agrees with the earlier result presented in Ref.~\cite{Barman:2023ymn}.
	
	For scatterings, the procedure is exactly the same, except that $\Gamma^{1\rightarrow 3}$ must be replaced by $\Gamma^{2\rightarrow 3}$ 
	and $m_\phi \to 2 m_\phi$ for annihilation
	in Eq.~\eqref{Eq:friedmann}.
	We then obtain
	\begin{equation}
	\oGW^{2\rightarrow 3}h^2 \simeq 10^{-19}\, \sigma^2\left(\frac{\Trh}{10^{10}~\text{GeV}}\right)^\frac{7}{3} \left(\frac{\rhoe}{6.25 \times 10^{62}~\text{GeV}^4}\right)^\frac{1}{6} \left(\frac{ 10^{13}~\text{GeV}}{m_\phi}\right)^2 \left(\frac{f}{10^8~\rm{Hz}}\right),
	\label{Eq:spectrumscattering}
	\end{equation}
	where $\rhoe$ represents the energy density stored in the inflaton field at the end of inflation.
	We normalized Eq.~\eqref{Eq:spectrumscattering} with parameters typical of the $T$-model $\alpha$-attractors, that is, $\rhoe \sim (5\times 10^{15})^4$~GeV$^4$ and $m_\phi \sim 10^{13}$~GeV. 
	For the sake of completeness, we report in Appendix~\ref{Sec:appgenericV} the complete expression summed over all the inflaton oscillatory modes for a generic potential $V(\phi)\propto \phi^n$.
	
	%%%%%%%%%%%%%%%%%%%%%%%%%%%%%%%%%%%%%%%%%%%%%%%%%%%%%%%%%%%%
	\begin{figure}
		\centering
		\includegraphics[scale=0.58]{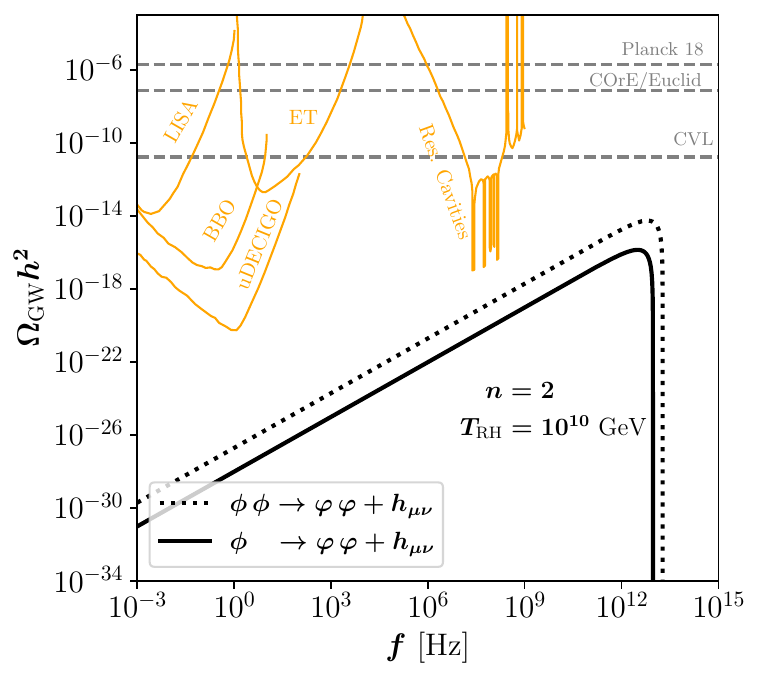}
		\includegraphics[scale=0.58]{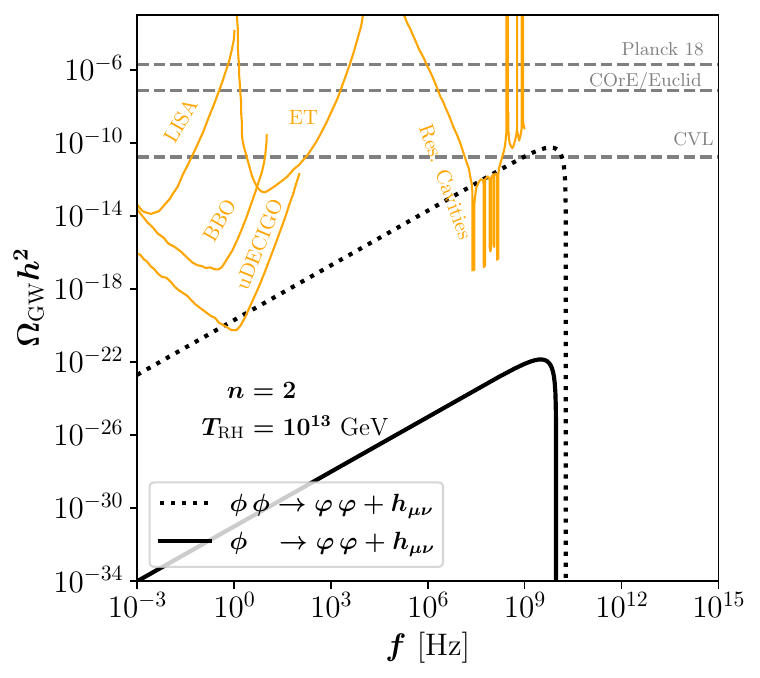}
		\includegraphics[scale=0.58]{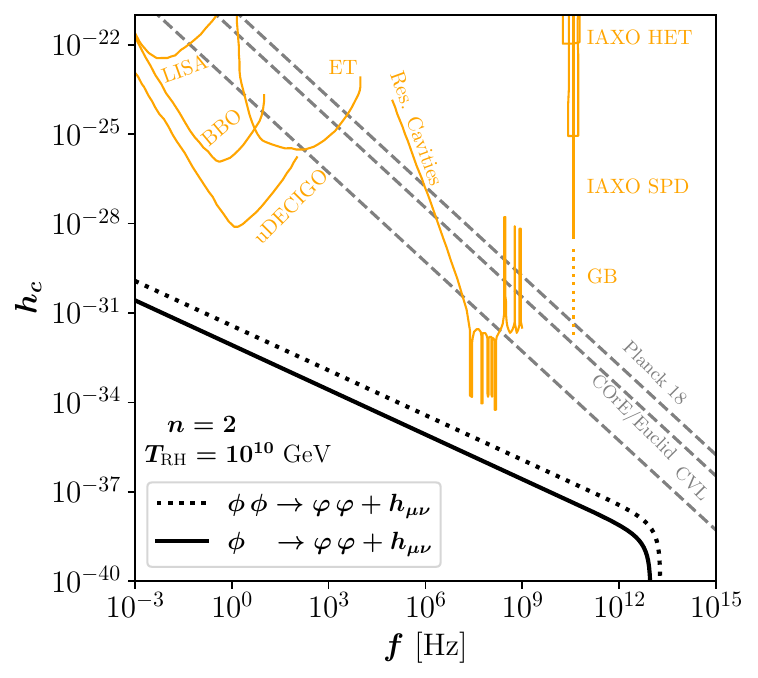}
		\includegraphics[scale=0.58]{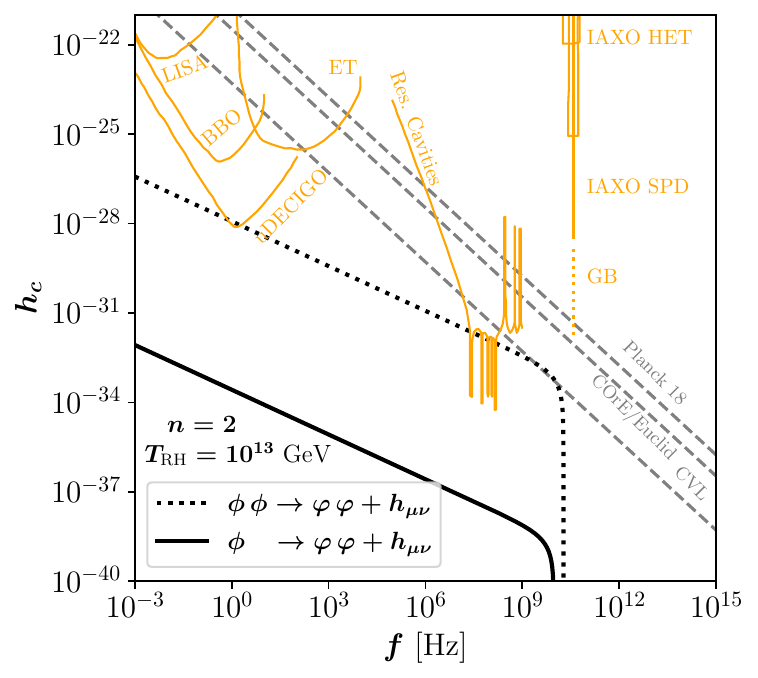}
		\caption{GW spectrum (top) and GW strain (bottom) for the quadratic inflaton potential $n=2$ with $\Trh = 10^{10}$~GeV (left) and $\Trh = 10^{13}$~GeV (right). Solid black lines represent the signal for inflaton decays with $\mu=10^{10}$~GeV whereas dashed black lines represent the signal for inflaton scattering with $\sigma = 1$. Sensitivities of future GW detectors are represented by solid orange lines.
		}
		\label{fig:GW_peak}
	\end{figure}
	%%%%%%%%%%%%%%%%%%%%%%%%%%%%%%%%%%%%%%%%%%%%%%%%%%%%%%%%%%%%
	We illustrate in the top left panel of Fig.~\ref{fig:GW_peak} the GW spectrum predicted for inflaton decay (solid black line) and scattering (dotted black line) assuming $\mu=10^{10}$~GeV, $\sigma=1$, $\Trh=10^{10}$~GeV and $n=2$. We note that the spectra (in the low-frequency regime) nicely match the approximate expressions in Eqs.~\eqref{Eq:spectrumdecay} and~\eqref{Eq:spectrumscattering}, respectively. As mentioned in the previous section, the emitted graviton energy from inflaton annihilation can be twice as large as that from the decay, and consequently, the upper limit of the GW frequency can be twice as large. This explains why the dotted curve (annihilation) peaks at a frequency twice as large as the solid line (decay).
	
	In addition, we overlay several proposed high-frequency GW detectors\footnote{See Ref.~\cite{Aggarwal:2020olq} for a recent review.} with high sensitivity in the figure. These include LISA~\cite{LISA:2017pwj}, the Einstein Telescope (ET)~\cite{Punturo:2010zz, Hild:2010id, Sathyaprakash:2012jk, Maggiore:2019uih}, the Big Bang Observer (BBO)~\cite{Crowder:2005nr, Corbin:2005ny, Harry:2006fi}, ultimate DECIGO (uDECIGO)~\cite{Seto:2001qf, Kudoh:2005as}, GW-electromagnetic wave conversion in vacuum (solid) and in a Gaussian beam (GB) (dotted)~\cite{Li:2003tv, Ringwald:2020ist}, resonant cavities~\cite{Berlin:2021txa, Berlin:2023grv, Herman:2022fau}, and the International Axion Observatory (IAXO)~\cite{Armengaud:2014gea, IAXO:2019mpb}. Sensitivity curves for these detectors were adapted from Refs.~\cite{Ringwald:2020ist, Ringwald:2022xif}. The energy stored in GWs exhibits characteristics similar to those of dark radiation, which, in turn, contributes to the effective number of neutrino species, denoted $N_{\text{eff}}$.\footnote{Note that the constraint from $N_{\text{eff}}$ on the amplitude of GWs, namely $\oGW h^2 \lesssim 5.6 \times 10^{-6} \DNeff$ only applies for the integrated energy density over logarithmic frequency. However, this is a good approximation over a broad range of frequencies and for the GW wavelength well inside the horizon at the moment when the constraint on $\DNeff$ is established~\cite{Caprini:2018mtu}.} In this paper, we present a compilation of various constraints related to this phenomenon. The Planck 2018 mission provides valuable information, establishing a 95\% CL result of $N_{\text{eff}} = 2.99 \pm 0.34$~\cite{Planck:2018vyg}. Future experiments, such as COrE~\cite{COrE:2011bfs} and Euclid~\cite{EUCLID:2011zbd}, are expected to significantly enhance these constraints at the $2\sigma$ level, resulting in $\DNeff \lesssim 0.013$. It is also interesting to mention a bound of $\DNeff \lesssim 3 \times 10^{-6}$ reported in Ref.~\cite{Ben-Dayan:2019gll} based on a hypothetical cosmic-variance-limited (CVL) CMB polarization experiment.
	We note that for a reheating temperature as low as $10^{10}$~GeV, it is not possible to detect graviton bremsstrahlung, even in future experiments.
	
	However, the top right panel of Fig.~\ref{fig:GW_peak} shows an equivalent result, except for $\Trh=10^{13}$~GeV, which corresponds roughly to an instantaneous reheating. We then clearly see that the spectrum derived from the scattering, in the case $n=2$, is larger by orders of magnitude than that generated by the inflaton decay. It is also interesting to note that there is a weak dependence of our results on the parameter $n$. We have made this explicit in Eq.~\eqref{Eq:genericn} in the case of scattering. Furthermore, in our analysis above, we distinguish between the reheating process, which involves $\Trh$, and the bremsstrahlung process. We can also ask whether it is possible to test what the reheating process is (decay or scattering) by measuring the graviton bremsstrahlung generated during the reheating period. But for that, one needs to look for $n \geq 4$, as reheating is not possible by scattering for $n=2$, which is the purpose of the following section.
	
	For the sake of completeness, and in order to project sensitivity curves from different GW experiments, it is also convenient to define the dimensionless strain $h_c$ in terms of the GW spectral energy density as~\cite{Maggiore:1999vm}
	\begin{equation}
	h_c(f) \equiv \frac{H_0}{f}\, \sqrt{\frac{3}{2 \pi^2}\, \oGW(f)} \simeq 1.26 \times 10^{-18} \left(\frac{\rm{Hz}}{f}\right) \sqrt{h^2\, \oGW(f)}\,,  
	\end{equation}
	where $H_0 \equiv H(T_0) \simeq 1.44 \times 10^{-42}$~GeV is the present-day Hubble parameter~\cite{Planck:2018vyg}. 
	As an illustration, in Fig.~\ref{fig:GW_peak} (bottom) we depict the strain parameter $h_c$ as a function of frequency $f$ for the same set of parameters as the one used in Fig.~\ref{fig:GW_peak} (top).
	
	%%%%%%%%%%%%%%%%%%%%%%%%%%%%%%%%%%%%%%%%%%%%%%%%%%%%%%%%%%%%
	\subsection{Combining graviton bremsstrahlung and reheating}
	%%%%%%%%%%%%%%%%%%%%%%%%%%%%%%%%%%%%%%%%%%%%%%%%%%%%%%%%%%%%
	In the previous section, the system of Boltzmann equations for the background was solved
	in the general case, without directly linking the reheating channel with the graviton emission channel.
	However, to compute the GW spectrum one has to keep track of the evolution of the differential GW energy density $\rGW$  and radiation $\rho_R$, and therefore instead of Eqs.~\eqref{eq:drPdt} and~\eqref{eq:drRdt} one has to solve~\cite{Barman:2023rpg}
	\begin{align}
	&\frac{d\rp}{dt} + \frac{6\, n}{2 + n}\, H\, \rp = - \frac{2\, n}{2 + n}\, \left(\gamma^{(0)} + \gamma^{(1)}\right), \label{eq:beq1}\\
	&\frac{d\rR}{dt} + 4\, H\, \rR = + \frac{2\, n}{2 + n}\, \gamma^{(0)} + \frac{2\, n}{2 + n}\, \int \frac{d\gamma^{(1)}}{d\Eom}\, \frac{E_\phi - \Eom}{E_\phi}\, d\Eom\,, \label{eq:beq2}\\
	&\frac{d\rGW}{dt} + 4\, H\, \rGW = + \frac{2\, n}{2 + n}\, \int \frac{d\gamma^{(1)}}{d\Eom}\, \frac{\Eom}{E_\phi}\, d\Eom\,, \label{eq:beq3}
	\end{align}
	where $\Eom$ corresponds to the graviton energy at the moment of production, and $E_\phi$ corresponds to the initial total inflaton energy, either $E_\phi = m_\phi$ for decays or $E_\phi = 2\, m_\phi$ for annihilations. The term $(E_\phi - \Eom)/ E_\phi$  describes the energy fraction that goes to radiation. The density rates $\gamma^{(0)}$ and $\gamma^{(1)}$ correspond to the cases without and with graviton emission, respectively. More specifically,  for inflaton decays, we have $\gamma^{(0)} =\Gamma^{1\to 2} \rp$ and $\gamma^{(1)} =\Gamma^{1\to 3} \rp$, and $\gamma^{(0)} =\Gamma^{2\to 2} \rp$ and $\gamma^{(1)} =\Gamma^{2\to 3} \rp$ for inflaton annihilations.\footnote{Note that if the decay final states have $N$ degrees of freedom, one shall multiply a factor $N$ in the rates.}
	
	For the GW spectrum, what we need is the differential GW energy density; therefore, it is convenient to write Eq.~\eqref{eq:beq3} as
	\begin{equation} \label{eq:dGWdR}
	\frac{d}{da} \frac{d(\rGW/\rR)}{d\Eom} \simeq \frac{2n}{2 + n} \frac{1}{a\, H(\Trh)} \left(\frac{\arh}{a}\right)^{\frac{3n}{2 + n} - 4 \alpha} \left[\frac{d\Gamma^{(1)}}{d\Eom} \frac{\Eom}{m_\phi} - \frac{d(\rGW/\rR)}{d\Eom} \Gamma^{(0)}\right].
	\end{equation}
	Note that $\Eom$ corresponds to the energy at the moment when gravitons are emitted. However, to compute the GW spectrum, we need to consider the redshift and sum over different (redshifted) energies. Therefore, in Eq.~\eqref{eq:dGWdR} the change of variable $\Eom(\Eom',\, a) = \Eom'\, \frac{\arh}{a}$, with $\Eom'$ being the energy at $a = \arh$, is required, leading to~\cite{Barman:2023rpg}
	\begin{equation} \label{eq:dGWdR2}
	\frac{d}{da} \frac{d(\rGW/\rR)}{d\Eom'} \simeq \frac{\arh}{a} \frac{2n}{2 + n} \frac{1}{a\, H(\Trh)} \left(\frac{\arh}{a}\right)^{\frac{3n}{2 + n} - 4 \alpha} \left[\frac{d\Gamma^{(1)}}{d\Eom'} \frac{\Eom'}{m_\phi} - \frac{a}{\arh} \frac{d(\rGW/\rR)}{d\Eom'} \Gamma^{(0)}\right],
	\end{equation}
	where the extra overall factor $\arh/a$ arises from the Jacobian related to the changing of variables.
	
	The solution for Eq.~\eqref{eq:dGWdR2} in the case of inflaton decays was recently reported in Ref.~\cite{Barman:2023rpg}. In this work, we focus on the yet unexplored case of inflaton annihilation. In that case, Eq.~\eqref{eq:dGWdR2} admits analytical solutions for different values of $n$. Recalling that reheating by inflaton annihilation is not viable in the case $n = 2$, we explore the scenarios with $n=4$ and $n=6$.
	\begin{itemize}
		\item For $n=4$, it follows that
		\begin{equation} \label{eq:ann_n=4}
		\frac{d(\rGW/\rR)}{d\Eom'} \simeq \frac{2\,\mrh}{ \, M_P^2 \, \pi^2} \left(1 - \frac{\Eom'}{\mrh}\right)^2 \left[\left(\frac{\Tmax}{\Trh}\right)^{8/9}-1\right],
		\end{equation}
		for $0 \leq \Eom' \leq \mrh$. 
		\item For $n=6$, one has
		\begin{align}\label{eq:ann_n=6}
		\frac{d(\rGW/\rR)}{d\Eom'} \simeq  \frac{3\,\mrh}{2 \pi^2 M_P^2 }  &\Bigg[ \left(\left(\frac{\Tmax}{\Trh}\right)^{16/9} - 1\right) - 4 \frac{\Eom'}{ \mrh} \left(\left(\frac{\Tmax}{\Trh}\right)^{8/9} - 1\right) \nonumber \\ 
		& \quad +\frac{16}{9} \left(\frac{\Eom'}{\mrh} \right)^2 \log\left(\frac{\Tmax}{\Trh}\right)\Bigg]
		\end{align}
		for $0 \leq \Eom' \leq \mrh$, while for $\mrh \leq \Eom' \leq \mrh \left(\frac{\Tmax}{\Trh}\right)^{8/9}$ it becomes
		\begin{align}
		\frac{d(\rGW/\rR)}{d\Eom'}  \simeq  \frac{3\, \mrh}{2\pi^2 \, M_P^2 } & \Bigg[ \left( \frac{\Tmax}{\Trh}\right)^{16/9} - 4 \frac{\Eom'}{\mrh} \left(\frac{\Tmax}{\Trh}\right)^{8/9} \nonumber \\ 
		& \quad + \left(\frac{\Eom'}{\mrh} \right)^2 \left(3 + 2 \log \left[\frac{\mrh}{\Eom'} \left(\frac{\Tmax}{\Trh}\right)^{8/9} \right]\right) \Bigg].
		\end{align}
	\end{itemize}
	Several comments on the regimes for $\Eom'$ shown above are in order. Note that the graviton energy at emission always lies in the range $0 < \Eom \leq m_\phi(a)$, which together with $\Eom' = \Eom \, a/\arh$ implies $0 < \Eom' \leq \mrh\, (a/\arh)^{2(n-4)/(n+2)}$. Subsequently, for $n = 4$, one has $0 < \Eom' \leq \mrh$. However, for $n = 6$, there are two energy regimes: low-energy gravitons in the range $0 < \Eom' \leq \mrh$ can be produced during the entire reheating period, while high-energy gravitons with $\mrh \leq \Eom' \leq \mrh\, \left(\Tmax/\Trh\right)^{8/9}$ can only be produced in the last stages of reheating. Additionally, we notice that the GW spectrum features a smaller boost proportional to powers of the ratio $\Tmax/\Trh$, compared to the decay case, as for annihilations the entropy dilution is more prominent, as shown in Fig.~\ref{fig:rho_T}.
	
	The primordial GW spectrum $\oGW(f)$ at present per logarithmic frequency $f$ is
	\begin{equation} \label{eq:oGW}
	\oGW(f) = \Omega_\gamma^{(0)}\, \frac{\gs(\Trh)}{\gs(T_0)} \left[\frac{\gss(T_0)}{\gss(\Trh)}\right]^{4/3}\,\frac{d(\rGW(\Trh)/\rR(\Trh))}{d\ln \Eom'}\,,
	\end{equation}
	where $\Omega_\gamma^{(0)} h^2 \simeq 2.47 \times 10^{-5}$ is the current photon density, $T_0 \simeq 2.73$~K is the CMB temperature~\cite{Planck:2018vyg}, and $\gss(T)$ is the number of relativistic degrees of freedom that contribute to the SM entropy.
	The present GW frequency is associated with the graviton energy $\Eom'$ at the end of reheating via 
	\begin{equation}
	f = \frac{\Eom'}{2 \pi}\, \frac{\arh}{a_0} = \frac{\Eom'}{2 \pi}\, \frac{T_0}{\Trh} \left[\frac{\gss(T_0)}{\gss(\Trh)}\right]^{1/3},
	\end{equation}
	considering the redshift of the graviton energy from the end of reheating and the present epoch. Note that the frequency is bounded from above since the graviton at production could carry at most half of the total inflaton energy, namely $\Eom(a) \leq m_\phi(a)$ for inflaton annihilation, which translates into
	\begin{equation} \label{eq:fup}
	f \leq \frac{\mrh}{2 \pi}\, \frac{\arh}{a_0} \left(\frac{\arh}{a}\right)^\frac{2 (n-4)}{n+2} \leq \frac{\mrh}{2 \pi}\, \frac{T_0}{\Trh} \left[\frac{\gss(T_0)}{\gss(\Trh)}\right]^{1/3} \times
	\begin{dcases}
	1 &\text{ for } n \leq 4\,,\\
	\left(\frac{\Tmax}{\Trh}\right)^\frac{2 (n-4)}{\alpha (n+2)} &\text{ for } n > 4\,.
	\end{dcases}
	\end{equation}
	In the case of an inflaton decay, $\Eom(a) \leq m_\phi(a)/2$, leading to an additional $1/2$ on the right-hand side of Eq.~\eqref{eq:fup}.
	
\begin{figure}[t!]
		\def\sepf{0.8}
		\centering
		\includegraphics[scale=\sepf]{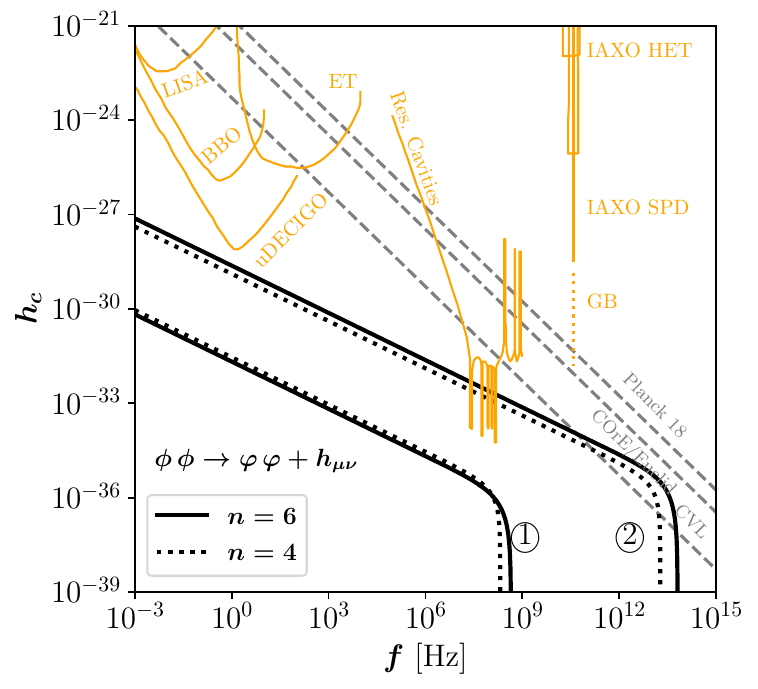}
		\caption{GW signature from graviton bremsstrahlung via inflaton annihilation during reheating, assuming the following parameters: \textcircled{1}  $\Trh = 10^{13}$~GeV,  $\mrh = 1.1 \times 10^{11}$~GeV, $\Tmax/\Trh = 43$ (for $n=4$) and $\Trh = 10^{13}$~GeV,  $\mrh = 2.6 \times 10^{10}$~GeV, $\Tmax/\Trh = 12$ (for $n=6$); \textcircled{2} $\mrh = 5 \times 10^{16}$~GeV, $\Trh = 5\times 10^{13}$~GeV and $\Tmax/\Trh = 4$ for both $n=4$ and $n=6$. }
		\label{fig:GW_ann}
	\end{figure}

	In Fig.~\ref{fig:GW_ann}, we present the GW spectrum for the benchmark parameters: \textcircled{1} $\Trh = 10^{13}$~GeV, $\mrh = 1.1 \times 10^{11}$~GeV, $\Tmax/\Trh = 43$ (for $n=4$) and $\Trh = 10^{13}$~GeV,  $\mrh = 2.6 \times 10^{10}$~GeV, $\Tmax/\Trh = 12$ (for $n=6$), and \textcircled{2} $\mrh = 5 \times 10^{16}$~GeV, $\Trh = 5\times 10^{13}$~GeV and $\Tmax/\Trh = 4$ for both $n=4$ and $n=6$. Note that the parameters in \textcircled{1} are derived from the $\alpha$-attractor $T$-model (cf. Fig.~\ref{fig:rho_T}). In addition, the model parameters we have considered satisfy the limits detailed in Section~\ref{sec:inflaton_ann}. GW signals are depicted for two different cases: $n=6$ (solid black line) and $n=4$ (dotted black line). For the case with $n=6$, the signal shows a boost factor $\propto \left(\Tmax/\Trh\right)^{16/9}$ (cf. Eq.~\eqref{eq:ann_n=6}), which exceeds the boost factor $\propto \left(\Tmax/\Trh\right)^{8/9}$ observed in the case $n=4$ (cf. Eq.~\eqref{eq:ann_n=4}), and therefore the strain of the GW spectrum corresponding to $n=6$ is expected to be higher. (Note that this conclusion is based on the same model parameters, that is, \textcircled{2}.)
	Furthermore, it should be noted that the upper limit of the GW frequency is higher for $n = 6$ compared to the case with $n = 4$, as explained by Eq.~\eqref{eq:fup}. The wider range of frequencies for $n = 6$ adds to the distinction between the two cases in the spectrum. These results suggest that the parameter space of the inflaton annihilation reheating scenario might be probed by next-generation GW detectors as gravitational cavities and $\DNeff$ experiments.
	%%%%%%%%%%%%%%%%%%%%%%%%%%%%%%%%%%%%%%%%%%%%%%%%%%%%%
	%%%%%%%%%%%%%%%%%%%%%%%%%%%%%%%%%%%%%%%%%%%%%%%%%%%%%
	
\begin{figure}[t!]
		\def\sepf{0.58}
		\centering
		\includegraphics[scale=\sepf]{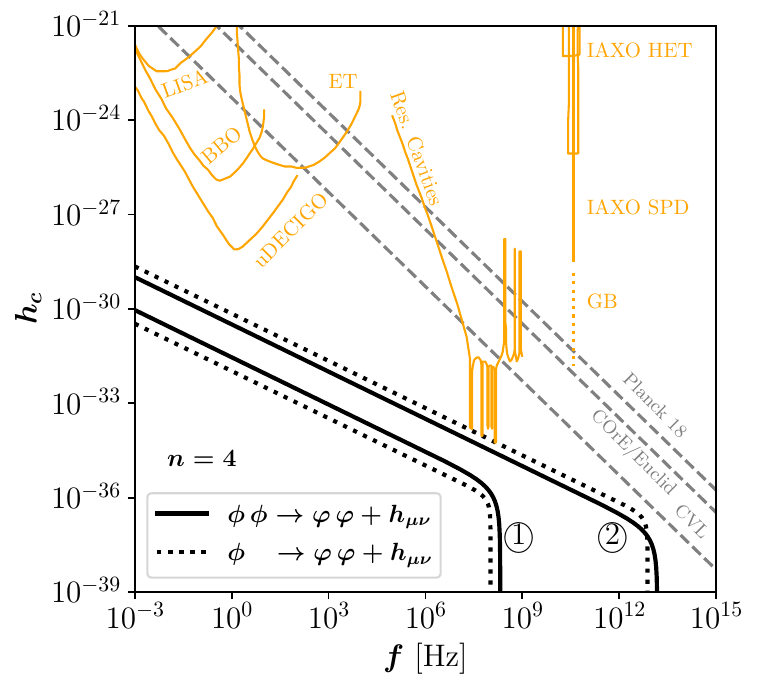}
		\includegraphics[scale=\sepf]{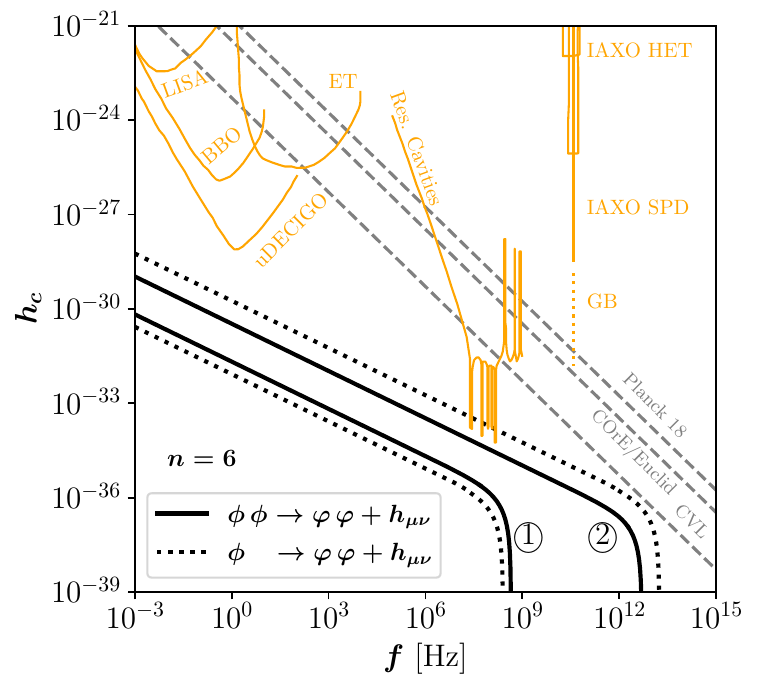}
		\caption{Comparison for GW signature from graviton bremsstrahlung via inflaton decay and annihilation during reheating. The left and right frames correspond to the spectrum with $n=4$ and $n=6$, respectively.}
		\label{fig:GW}
	\end{figure}

	To facilitate comparison of our results on inflaton annihilation with inflaton decays, in Fig.~\ref{fig:GW}, we show the GW spectrum of several benchmark parameters. In the left panel with $n=4$, the parameters considered are: \textcircled{1} $\Trh = 10^{13}$~GeV, $\mrh = 1.1 \times 10^{11}$~GeV, $\Tmax/\Trh = 3.6$  (for decay) and $\Trh = 10^{13}$~GeV,  $\mrh = 1.1 \times 10^{11}$~GeV, $\Tmax/\Trh = 43$ (for annihilation); \textcircled{2} $\Trh = 10^{12}$~GeV, $\Tmax/\Trh = 7$, $\mrh = 8 \times 10^{14}$~GeV for both decay and annihilation. In the right frame with $n=6$: \textcircled{1} $\Trh = 10^{13}$~GeV, $\mrh = 2.6 \times 10^{10}$~GeV, $\Tmax/\Trh = 2.4$  (for decay) and $\Trh = 10^{13}$~GeV,  $\mrh = 2.6 \times 10^{10}$~GeV, $\Tmax/\Trh = 12$ (for annihilation);   \textcircled{2} $\Trh =3 \times 10^{12}$~GeV, $\Tmax/\Trh = 3$, $\mrh = 3 \times 10^{14}$~GeV for both decay and annihilation. 
	
	Solid lines correspond to the case where the inflaton annihilates, while dotted curves correspond to the case of decay.  For $n = 4$, we notice that the GW frequency in the annihilation scenario can be shifted to a higher value compared to the decay model; this is because the maximum energy of the emitted graviton can be twice higher in the annihilation scenario. In the context of the same model parameters, e.g., \textcircled{2}, we find that the GW amplitude in the case of annihilation is smaller than that in the case of decay. This is because for annihilation, the spectrum has an enhancement during reheating $\propto \left(\Tmax/\Trh\right)^{8/9}$ (cf. Eq.~\eqref{eq:ann_n=4}) that is lower than that in the case of decays, where the enhancement is $\propto \left(\Tmax/\Trh\right)^{8/3}$~\cite{Barman:2023rpg}. This distinctive enhancement originates from the evolution of the temperature and hence the dilution effect during reheating. Due to the steeper slope of temperature evolution in the annihilation scenario (cf. Fig.~\ref{fig:rho_T}), it is expected that for larger $n$, the dilution effect becomes more prominent. Consequently, the spectrum in the annihilation model becomes suppressed compared to that in the decay case, as shown in the figure.
	With a progressive increase in the parameter $n$, the distinction between the GW spectra in the two scenarios considered becomes increasingly pronounced, as illustrated in the right panel.
	%%%%%%%%%%%%%%%%%%%%%%%%%%%%%%%%%%%%%%%%%%%%%%%
	%%%%%%%%%%%%%%%%%%%%%%%%%%%%%%%%%%%%%%%%%%%%%%%
	For the aforementioned benchmark parameters, we notice that the GW signature might be within the sensitivity of future cavity experiments.
	Specifically, for $n=6$, we observe that the GW spectrum exhibits a discernible signal that may be detectable by resonant cavity detectors. On the contrary, the amplitude of the GW spectrum in the annihilation scenario is considerably lower and lies beyond the range of sensitivity of the cavity detectors. 
	
	In the event that a null signal is observed in future high-frequency GW measurements, it would enable the exclusion of certain regions of the parameter space.
	Consequently, our proposed approach, utilizing bremsstrahlung-induced GWs, could give useful information on the reheating dynamics.
	
	%%%%%%%%%%%%%%%%%%%%%%%%%%%%%%%%%%%%%%%%%%%%%%%%%%%%%%%%%%
	\section{Conclusions} \label{sec:concl}
	%%%%%%%%%%%%%%%%%%%%%%%%%%%%%%%%%%%%%%%%%%%%%%%%%%%%%%%%%%
	In this study, we have revisited graviton bremsstrahlung during the phase of inflationary reheating in the presence of an inflaton field $\phi$ oscillating around a generic monomial potential $\phi^n$. Unlike previous analyses, we have examined for the first time the gravitational wave (GW) spectrum within the context of a reheating scenario involving inflaton annihilations. We calculated the corresponding graviton emission rate, which is presented in Eq.~\eqref{eq:2to3diffrate} and the corresponding GW spectra are illustrated in Fig.~\ref{fig:GW_peak}. This is complementary to previous studies where only inflaton decays were assumed~\cite{Nakayama:2018ptw, Huang:2019lgd, Ghoshal:2022kqp, Barman:2023ymn, Barman:2023rpg, Kanemura:2023pnv}.
	
	We showed first that, depending on the reheating temperature, in the case of a quadratic potential with $n=2$, the scattering of the inflaton can give a much larger amount of graviton through bremsstrahlung than the decay channel, as one can see in the right panel of Fig.~\ref{fig:GW_peak}. Interestingly, these spectra could be in the region probed by next-generation GW observatories.
	
	We then concentrated our study on the assumption that the channel responsible for the bremsstrahlung is also responsible for the reheating process. The reheating temperature is then directly related to the GW spectrum. For $n>2$ we found that the GW spectrum features a power-law enhancement on the ratio $\Tmax/\Trh$, whose effect becomes more prominent with increasing $n$, as shown in Fig.~\ref{fig:GW_ann}. Compared to the inflaton decay scenario, we found that the GW spectrum exhibits distinct characteristics, depicted in Fig.~\ref{fig:GW}, which could potentially allow us to distinguish between decays and annihilations. These distinctions arise from the disparate scaling behaviors exhibited by the radiation released during the reheating process (cf. Fig.~\ref{fig:rho_T}), which behaves as a dilution effect for the GW spectrum. Particularly noteworthy is the fact that, for certain model parameters, the GW spectrum in the annihilation scenario is smaller than in the decay case, where the latter could fall within the detectability range of future cavity detectors. 
	Our results thus suggest a new potential avenue for probing the parameter space of reheating by utilizing bremsstrahlung-induced GWs.
	
	%%%%%%%%%%%%%%%%%%%%%%%%%%%%%%%%%%%%%%%%%%%
	\section*{Acknowledgments}
	%%%%%%%%%%%%%%%%%%%%%%%%%%%%%%%%%%%%%%%%%%
	The authors want to thank Basabendu Barman, Mathias Pierre and  Óscar Zapata for very fruitful discussions.
	N.B. received funding from the Spanish FEDER / MCIU-AEI under the grant FPA2017-84543-P and thanks the ICTP staff for their hospitality during his stay when this study was completed.
	S.C. and Y.M. acknowledge the support of the Institut Pascal at Université Paris-Saclay during the Paris-Saclay Astroparticle Symposium 2023, with the support of the P2IO Laboratory of Excellence (program ``Investissements d'avenir'' ANR-11-IDEX-0003-01 Paris-Saclay and ANR-10-LABX-0038), the P2I axis of the Graduate School of Physics of Université Paris-Saclay, as well as IJCLab, CEA, IAS, OSUPS, and the IN2P3 master project UCMN. 
	Y.X. has received support from the Cluster of Excellence ``Precision Physics, Fundamental Interactions, and Structure of Matter'' (PRISMA$^+$ EXC 2118/1) funded by the Deutsche Forschungsgemeinschaft (DFG, German Research Foundation) within the German Excellence Strategy (Project No. 390831469).
	This project has received funding and support from the European Union's Horizon 2020 research and innovation program under the Marie Sklodowska-Curie grant agreement No. 860881-HIDDeN.
	
	%%%%%%%%%%%%%%%%%%%%%%%%%%%%%%%%%%%%%%%%%%%%%%%%%%%%%%%%%%
	\appendix
	%%%%%%%%%%%%%%%%%%%%%%%%%%%%%%%%%%%%%%%%%%%%%%%%%%%%%%%%%%
	
	%%%%%%%%%%%%%%%%%%%%%%%%%%%%%%%%%%%%%%%%%%%%%%%%%%%%%%%%%%
	\section{Graviton Bremsstrahlung for a Monomial Potential}
	\label{Sec:appgenericV}
	%%%%%%%%%%%%%%%%%%%%%%%%%%%%%%%%%%%%%%%%%%%%%%%%%%%%%%%%%%
	We can solve Eq.~\eqref{eq:drhoGWdE} for a generic potential of the inflaton during reheating $V(\phi)=\lambda\, M_P^4 \left(\frac{\phi}{M_P}\right)^n$. We obtain the generic solution for the $1\rightarrow 3 $ process
	\beq
	\left.\frac{d\rGW}{d\Eom}\right|_{\arh}^{1\rightarrow3} = \frac{\sqrt{3} \mu^2 M_P}{32\pi^3} \alpha_n^2 \left(\frac{\rhorh}{M_P^4}\right)^\frac12 \int_{\aend}^{\arh} \frac{da}{\arh} \left(\frac{a}{\arh}\right)^{\frac{6}{n+2}} \sum\limits_{j=1}^{\infty} j^2 |\mathcal{P}_j|^2 \left(1 - 2 x_j(a)\right)^2,
	\eeq
	for $y \to 0$.
	For the case $n=2$, the frequency of the oscillation is fixed as $\varpi=m_\phi$, and there is only one mode in the Fourier decomposition corresponding to $j=1$ associated with the coefficient $\mathcal{P}_1=\frac{1}{2}$. Hence, ignoring the redshift dependence of $x_j$,\footnote{For a complete treatment of the energy redshift during reheating we refer the reader to Ref.~\cite{Barman:2023rpg}.} we recover
	\beq
	\left.\frac{d\rGW}{d\Eom}\right|_{\arh}^{1\rightarrow3}=\frac{\sqrt{3}\mu^2M_P}{160\pi^3}\left(\frac{\rhorh}{M_P^4}\right)^{1/2}\left[1-\left(\frac{\aend}{\arh}\right)^{5/2}\right]\left(1-2\frac{E_w}{m_\phi}\right)^2.
	\eeq
	The relic density of GWs is obtained by
	\beq
	\Omega^0_{\rm GW} h^2 = \frac{h^2}{\rho_c^0}\left.\frac{d\rGW}{d\Eom}\right|_{\arh}\left(\frac{\arh}{a_0}\right)^4\times \Eom(\arh)\,,
	\eeq
	where we can relate the present energy of the gravitons with the current frequency of the GW, $\Eom(\arh)= 2\pi\, f\, (a_0/\arh)$. This leads in case $n=2$ to the following relic density for $\aend \ll \arh$ and $\Eom(\arh) \ll m_\phi$
	\beq
	\Omega^0_{\rm GW} h^2 \simeq \frac{h^2\mu^2\sqrt{3}}{80\pi^2} \times \frac{f}{M_P} \times \frac{\Omega_R^0}{(\rhorh\rho_R^0)^{1/4}}\,. 
	\eeq
	For the generic case $n>2$, the integration of the energy density spectrum is generally more involved, as the energy in each mode is time dependent, leading to mode-dependent boundaries of the integral. We give an approximate solution for a part of the spectrum where $\Eom(\arh) \ll \omega(\arh)$. In this case, all the modes contribute similarly to the production of gravitons until the end of reheating, with a contribution proportional to $j^2|\mathcal{P}_j|^2$ for each mode $j$ 
	\beq
	\left.\frac{d\rGW}{d\Eom}\right|_{\arh}^{1\rightarrow3}\simeq \frac{\sqrt{3}\mu^2M_P}{32\pi^3}\alpha_n^2\left(\frac{\rhorh}{M_P^4}\right)^{1/2} \left(\frac{n+2}{n+8}\right)\left[1-\left(\frac{\aend}{\arh}\right)^{\frac{n+8}{n+2}}\right]\sum\limits_{j=1}^{\infty}j^2|\mathcal{P}_j|^2
	\eeq
	Hence, we see that in the low-frequency limit, the influence of the shape of the inflaton potential is only a change in the amplitude of the spectrum
	\beq
	\frac{\Omega_{\rm GW}^{n>2}}{\Omega_{\rm GW}^{n=2}} \simeq \frac{n+2}{n+8}\frac{\alpha_n^2}{5}\sum\limits_{j=1}^{\infty}j^2|\mathcal{P}_j|^2\,.
	\eeq
	Computing the integral with all modes provides a peak structure in the GW spectrum that we did not consider in our analysis.
	
	Now, we consider the scattering channel. We can obtain similar expressions for the $2\rightarrow3$ process
	\beq
	\left.\frac{d\rGW}{d\Eom}\right|_{\arh}^{2\to 3} = \frac{\sqrt{3} \sigma^2 M_P^3}{8\pi^3 \lambda^{2/n}} \alpha_n^2 \left(\frac{\rhorh}{M_P^4}\right)^{\frac{n+4}{2n}} \int_{\aend}^{\arh} \frac{da}{\arh}\left(\frac{a}{\arh}\right)^{-\frac{6}{n+2}} \sum\limits_{j=1}^{\infty}j^2 |\mathcal{P}_j^{(2)}|^2 \left(1-x_j(a)\right)^2.
	\eeq
	In the case $n=2$, we recover
	\beq
	\left.\frac{d\rGW}{d\Eom}\right|_{\arh}^{2\rightarrow3}=\frac{\sqrt{3}\sigma^2M_P^3}{16\pi^3\lambda}\left(\frac{\rhorh}{M_P^4}\right)^{3/2} \left[\left(\frac{\arh}{\aend}\right)^{1/2} -1\right]\left(1-\frac{\Eom}{m_\phi}\right)^2,
	\eeq
	leading, in the limit $\aend \ll \arh $ and $\Eom(\arh) \ll m_\phi$, to
	\beq
	\Omega^0_{\rm GW} h^2 \simeq \frac{h^2\sigma^2\sqrt{3}}{4\pi^2} \times \frac{f}{M_P} \times \frac{\rhoe^{1/6}\rhorh^{7/12}\Omega_R^0}{m_\phi^2(\rho_R^0)^{1/4}}\,,
	\eeq
	where we have used the relation $\lambda = m_\phi^2 / (2M_P^2)$ considering the quadratic potential of the inflaton.
	
	In the general case $n>2$, again the situation is more complicated as the frequency of the background oscillations evolves with time. However, we give an approximate solution for a part of the spectrum where $\Eom(\arh)\ll \omega(\arh)$
	\begin{equation}
	\left.\frac{d\rGW}{d\Eom}\right|_{\arh}^{2\rightarrow3} \simeq \frac{\sqrt{3} \sigma^2M_P^3}{8\pi^3\lambda^{1/2} } \alpha_4^2 \left(\frac{\rhorh}{M_P^4}\right) \ln{\left(\frac{\rhoe}{\rhorh}\right)} \sum \limits_{j=1}^{\infty} j^2 \left|\mathcal{P}_j^{(2)}\right|^2
	\end{equation}
	for $n=4$, or
	\begin{equation}
	\left.\frac{d\rGW}{d\Eom}\right|_{\arh}^{2\rightarrow3} \simeq \frac{n+2}{n-4} \frac{\sqrt{3} \sigma^2 M_P^3}{8 \pi^3\lambda^{2/n}} \alpha_n^2 \left(\frac{\rhorh}{M_P^4}\right)^{\frac{n+4}{2n}} \left[1 - \left(\frac{\aend}{\arh}\right)^{\frac{n-4}{n+2}}\right] \sum \limits_{j=1}^{\infty} j^2 \left|\mathcal{P}_j^{(2)}\right|^2
	\end{equation}
	for $n>4$.
	In the low-frequency limit, the influence of the shape of the inflaton potential for the $2\to3$ process again is only a change in the amplitude of the spectrum
	\begin{align}
	\frac{\Omega_{\rm GW}^{n=4}}{\Omega_{\rm GW}^{n=2}} &\simeq 2 \alpha_4^2 \lambda^{\frac{1}{2}}\sum\limits_{j=1}^{\infty}j^2 \left|\mathcal{P}^{(2)}_j\right|^2 ~~~~~~~~~~~~ (n=4),\\
	\frac{\Omega_{\rm GW}^{n>2}}{\Omega_{\rm GW}^{n=2}} &\simeq \frac{n+2}{n-4} \frac{2 \alpha_n^2}{\lambda^{\frac{2-n}{n}}} \sum\limits_{j=1}^{\infty}j^2 \left|\mathcal{P}^{(2)}_j\right|^2 ~~~~~~ (n>4).
	\label{Eq:genericn}
	\end{align}
	
	%%%%%%%%%%%%%%%%%%%%%%%%%%%%%%%%%%%%%%%%%%
	\bibliographystyle{JHEP}
	\bibliography{biblio}

\providecommand{\href}[2]{#2}\begingroup\raggedright\begin{thebibliography}{10}

\bibitem{Lyth:2009zz}
D.H.~Lyth and A.R.~Liddle, \emph{{The primordial density perturbation:
  Cosmology, inflation and the origin of structure}} (2009).

\bibitem{Kofman:1997yn}
L.~Kofman, A.D.~Linde and A.A.~Starobinsky, \emph{{Towards the theory of
  reheating after inflation}},
  \href{https://doi.org/10.1103/PhysRevD.56.3258}{\emph{Phys. Rev. D}
  {\bfseries 56} (1997) 3258}
  [\href{https://arxiv.org/abs/hep-ph/9704452}{{\ttfamily hep-ph/9704452}}].

\bibitem{Weinberg:1965nx}
S.~Weinberg, \emph{{Infrared photons and gravitons}},
  \href{https://doi.org/10.1103/PhysRev.140.B516}{\emph{Phys. Rev.} {\bfseries
  140} (1965) B516}.

\bibitem{Barker:1969jk}
B.M.~Barker, S.N.~Gupta and J.~Kaskas, \emph{{Graviton bremsstrahlung and
  infrared divergence}},
  \href{https://doi.org/10.1103/PhysRev.182.1391}{\emph{Phys. Rev.} {\bfseries
  182} (1969) 1391}.

\bibitem{Ghiglieri:2015nfa}
J.~Ghiglieri and M.~Laine, \emph{{Gravitational wave background from Standard
  Model physics: Qualitative features}},
  \href{https://doi.org/10.1088/1475-7516/2015/07/022}{\emph{JCAP} {\bfseries
  07} (2015) 022} [\href{https://arxiv.org/abs/1504.02569}{{\ttfamily
  1504.02569}}].

\bibitem{Ghiglieri:2020mhm}
J.~Ghiglieri, G.~Jackson, M.~Laine and Y.~Zhu, \emph{{Gravitational wave
  background from Standard Model physics: Complete leading order}},
  \href{https://doi.org/10.1007/JHEP07(2020)092}{\emph{JHEP} {\bfseries 07}
  (2020) 092} [\href{https://arxiv.org/abs/2004.11392}{{\ttfamily
  2004.11392}}].

\bibitem{Ringwald:2020ist}
A.~Ringwald, J.~Sch\"utte-Engel and C.~Tamarit, \emph{{Gravitational Waves as a
  Big Bang Thermometer}},
  \href{https://doi.org/10.1088/1475-7516/2021/03/054}{\emph{JCAP} {\bfseries
  03} (2021) 054} [\href{https://arxiv.org/abs/2011.04731}{{\ttfamily
  2011.04731}}].

\bibitem{Klose:2022knn}
P.~Klose, M.~Laine and S.~Procacci, \emph{{Gravitational wave background from
  non-Abelian reheating after axion-like inflation}},
  \href{https://doi.org/10.1088/1475-7516/2022/05/021}{\emph{JCAP} {\bfseries
  05} (2022) 021} [\href{https://arxiv.org/abs/2201.02317}{{\ttfamily
  2201.02317}}].

\bibitem{Ringwald:2022xif}
A.~Ringwald and C.~Tamarit, \emph{{Revealing the cosmic history with
  gravitational waves}},
  \href{https://doi.org/10.1103/PhysRevD.106.063027}{\emph{Phys. Rev. D}
  {\bfseries 106} (2022) 063027}
  [\href{https://arxiv.org/abs/2203.00621}{{\ttfamily 2203.00621}}].

\bibitem{Ghiglieri:2022rfp}
J.~Ghiglieri, J.~Sch\"utte-Engel and E.~Speranza, \emph{{Freezing-In
  Gravitational Waves}},  \href{https://arxiv.org/abs/2211.16513}{{\ttfamily
  2211.16513}}.

\bibitem{Nakayama:2018ptw}
K.~Nakayama and Y.~Tang, \emph{{Stochastic Gravitational Waves from Particle
  Origin}}, \href{https://doi.org/10.1016/j.physletb.2018.11.023}{\emph{Phys.
  Lett. B} {\bfseries 788} (2019) 341}
  [\href{https://arxiv.org/abs/1810.04975}{{\ttfamily 1810.04975}}].

\bibitem{Huang:2019lgd}
D.~Huang and L.~Yin, \emph{{Stochastic Gravitational Waves from Inflaton
  Decays}}, \href{https://doi.org/10.1103/PhysRevD.100.043538}{\emph{Phys. Rev.
  D} {\bfseries 100} (2019) 043538}
  [\href{https://arxiv.org/abs/1905.08510}{{\ttfamily 1905.08510}}].

\bibitem{Ghoshal:2022kqp}
A.~Ghoshal, R.~Samanta and G.~White, \emph{{Bremsstrahlung high-frequency
  gravitational wave signatures of high-scale nonthermal leptogenesis}},
  \href{https://doi.org/10.1103/PhysRevD.108.035019}{\emph{Phys. Rev. D}
  {\bfseries 108} (2023) 035019}
  [\href{https://arxiv.org/abs/2211.10433}{{\ttfamily 2211.10433}}].

\bibitem{Barman:2023ymn}
B.~Barman, N.~Bernal, Y.~Xu and {\'O}.~Zapata, \emph{{Gravitational wave from
  graviton Bremsstrahlung during reheating}},
  \href{https://doi.org/10.1088/1475-7516/2023/05/019}{\emph{JCAP} {\bfseries
  05} (2023) 019} [\href{https://arxiv.org/abs/2301.11345}{{\ttfamily
  2301.11345}}].

\bibitem{Barman:2023rpg}
B.~Barman, N.~Bernal, Y.~Xu and {\'O}.~Zapata, \emph{{Bremsstrahlung-induced
  gravitational waves in monomial potentials during reheating}},
  \href{https://doi.org/10.1103/PhysRevD.108.083524}{\emph{Phys. Rev. D}
  {\bfseries 108} (2023) 083524}
  [\href{https://arxiv.org/abs/2305.16388}{{\ttfamily 2305.16388}}].

\bibitem{Kanemura:2023pnv}
S.~Kanemura and K.~Kaneta, \emph{{Gravitational Waves from Particle Decays
  during Reheating}},  \href{https://arxiv.org/abs/2310.12023}{{\ttfamily
  2310.12023}}.

\bibitem{Garcia:2020wiy}
M.A.G.~Garcia, K.~Kaneta, Y.~Mambrini and K.A.~Olive, \emph{{Inflaton
  Oscillations and Post-Inflationary Reheating}},
  \href{https://doi.org/10.1088/1475-7516/2021/04/012}{\emph{JCAP} {\bfseries
  04} (2021) 012} [\href{https://arxiv.org/abs/2012.10756}{{\ttfamily
  2012.10756}}].

\bibitem{Garcia:2020eof}
M.A.G.~Garcia, K.~Kaneta, Y.~Mambrini and K.A.~Olive, \emph{{Reheating and
  Post-inflationary Production of Dark Matter}},
  \href{https://doi.org/10.1103/PhysRevD.101.123507}{\emph{Phys. Rev. D}
  {\bfseries 101} (2020) 123507}
  [\href{https://arxiv.org/abs/2004.08404}{{\ttfamily 2004.08404}}].

\bibitem{Barman:2021ugy}
B.~Barman and N.~Bernal, \emph{{Gravitational SIMPs}},
  \href{https://doi.org/10.1088/1475-7516/2021/06/011}{\emph{JCAP} {\bfseries
  06} (2021) 011} [\href{https://arxiv.org/abs/2104.10699}{{\ttfamily
  2104.10699}}].

\bibitem{Choi:1994ax}
S.Y.~Choi, J.S.~Shim and H.S.~Song, \emph{{Factorization and polarization in
  linearized gravity}},
  \href{https://doi.org/10.1103/PhysRevD.51.2751}{\emph{Phys. Rev. D}
  {\bfseries 51} (1995) 2751}
  [\href{https://arxiv.org/abs/hep-th/9411092}{{\ttfamily hep-th/9411092}}].

\bibitem{Turner:1983he}
M.S.~Turner, \emph{{Coherent Scalar Field Oscillations in an Expanding
  Universe}}, \href{https://doi.org/10.1103/PhysRevD.28.1243}{\emph{Phys. Rev.
  D} {\bfseries 28} (1983) 1243}.

\bibitem{Bernal:2022wck}
N.~Bernal and Y.~Xu, \emph{{WIMPs during reheating}},
  \href{https://doi.org/10.1088/1475-7516/2022/12/017}{\emph{JCAP} {\bfseries
  12} (2022) 017} [\href{https://arxiv.org/abs/2209.07546}{{\ttfamily
  2209.07546}}].

\bibitem{Starobinsky:1980te}
A.A.~Starobinsky, \emph{{A New Type of Isotropic Cosmological Models Without
  Singularity}},
  \href{https://doi.org/10.1016/0370-2693(80)90670-X}{\emph{Phys. Lett. B}
  {\bfseries 91} (1980) 99}.

\bibitem{Drees:2021wgd}
M.~Drees and Y.~Xu, \emph{{Small field polynomial inflation: reheating,
  radiative stability and lower bound}},
  \href{https://doi.org/10.1088/1475-7516/2021/09/012}{\emph{JCAP} {\bfseries
  09} (2021) 012} [\href{https://arxiv.org/abs/2104.03977}{{\ttfamily
  2104.03977}}].

\bibitem{Bernal:2021qrl}
N.~Bernal and Y.~Xu, \emph{{Polynomial inflation and dark matter}},
  \href{https://doi.org/10.1140/epjc/s10052-021-09694-5}{\emph{Eur. Phys. J. C}
  {\bfseries 81} (2021) 877}
  [\href{https://arxiv.org/abs/2106.03950}{{\ttfamily 2106.03950}}].

\bibitem{Drees:2022aea}
M.~Drees and Y.~Xu, \emph{{Large field polynomial inflation: parameter space,
  predictions and (double) eternal nature}},
  \href{https://doi.org/10.1088/1475-7516/2022/12/005}{\emph{JCAP} {\bfseries
  12} (2022) 005} [\href{https://arxiv.org/abs/2209.07545}{{\ttfamily
  2209.07545}}].

\bibitem{Kallosh:2013hoa}
R.~Kallosh and A.~Linde, \emph{{Universality Class in Conformal Inflation}},
  \href{https://doi.org/10.1088/1475-7516/2013/07/002}{\emph{JCAP} {\bfseries
  07} (2013) 002} [\href{https://arxiv.org/abs/1306.5220}{{\ttfamily
  1306.5220}}].

\bibitem{Kallosh:2013yoa}
R.~Kallosh, A.~Linde and D.~Roest, \emph{{Superconformal Inflationary
  $\alpha$-Attractors}},
  \href{https://doi.org/10.1007/JHEP11(2013)198}{\emph{JHEP} {\bfseries 11}
  (2013) 198} [\href{https://arxiv.org/abs/1311.0472}{{\ttfamily 1311.0472}}].

\bibitem{Mukaida:2015ria}
K.~Mukaida and M.~Yamada, \emph{{Thermalization Process after Inflation and
  Effective Potential of Scalar Field}},
  \href{https://doi.org/10.1088/1475-7516/2016/02/003}{\emph{JCAP} {\bfseries
  02} (2016) 003} [\href{https://arxiv.org/abs/1506.07661}{{\ttfamily
  1506.07661}}].

\bibitem{Garcia:2018wtq}
M.A.G.~Garcia and M.A.~Amin, \emph{{Prethermalization production of dark
  matter}}, \href{https://doi.org/10.1103/PhysRevD.98.103504}{\emph{Phys. Rev.
  D} {\bfseries 98} (2018) 103504}
  [\href{https://arxiv.org/abs/1806.01865}{{\ttfamily 1806.01865}}].

\bibitem{Chowdhury:2023jft}
D.~Chowdhury and A.~Hait, \emph{{Thermalization in the presence of a
  time-dependent dissipation and its impact on dark matter production}},
  \href{https://doi.org/10.1007/JHEP09(2023)085}{\emph{JHEP} {\bfseries 09}
  (2023) 085} [\href{https://arxiv.org/abs/2302.06654}{{\ttfamily
  2302.06654}}].

\bibitem{Amin:2014eta}
M.A.~Amin, M.P.~Hertzberg, D.I.~Kaiser and J.~Karouby, \emph{{Nonperturbative
  Dynamics Of Reheating After Inflation: A Review}},
  \href{https://doi.org/10.1142/S0218271815300037}{\emph{Int. J. Mod. Phys. D}
  {\bfseries 24} (2014) 1530003}
  [\href{https://arxiv.org/abs/1410.3808}{{\ttfamily 1410.3808}}].

\bibitem{Garcia:2021iag}
M.A.G.~Garcia, K.~Kaneta, Y.~Mambrini, K.A.~Olive and S.~Verner,
  \emph{{Freeze-in from preheating}},
  \href{https://doi.org/10.1088/1475-7516/2022/03/016}{\emph{JCAP} {\bfseries
  03} (2022) 016} [\href{https://arxiv.org/abs/2109.13280}{{\ttfamily
  2109.13280}}].

\bibitem{Lozanov:2016hid}
K.D.~Lozanov and M.A.~Amin, \emph{{Equation of State and Duration to Radiation
  Domination after Inflation}},
  \href{https://doi.org/10.1103/PhysRevLett.119.061301}{\emph{Phys. Rev. Lett.}
  {\bfseries 119} (2017) 061301}
  [\href{https://arxiv.org/abs/1608.01213}{{\ttfamily 1608.01213}}].

\bibitem{Garcia:2023eol}
M.A.G.~Garcia and M.~Pierre, \emph{{Reheating after inflaton fragmentation}},
  \href{https://doi.org/10.1088/1475-7516/2023/11/004}{\emph{JCAP} {\bfseries
  11} (2023) 004} [\href{https://arxiv.org/abs/2306.08038}{{\ttfamily
  2306.08038}}].

\bibitem{Garcia:2023dyf}
M.A.G.~Garcia, M.~Gross, Y.~Mambrini, K.A.~Olive, M.~Pierre and J.-H.~Yoon,
  \emph{{Effects of fragmentation on post-inflationary reheating}},
  \href{https://doi.org/10.1088/1475-7516/2023/12/028}{\emph{JCAP} {\bfseries
  12} (2023) 028} [\href{https://arxiv.org/abs/2308.16231}{{\ttfamily
  2308.16231}}].

\bibitem{Bernal:2020qyu}
N.~Bernal, J.~Rubio and H.~Veerm\"ae, \emph{{UV Freeze-in in Starobinsky
  Inflation}}, \href{https://doi.org/10.1088/1475-7516/2020/10/021}{\emph{JCAP}
  {\bfseries 10} (2020) 021}
  [\href{https://arxiv.org/abs/2006.02442}{{\ttfamily 2006.02442}}].

\bibitem{Clery:2021bwz}
S.~Clery, Y.~Mambrini, K.A.~Olive and S.~Verner, \emph{{Gravitational portals
  in the early Universe}},
  \href{https://doi.org/10.1103/PhysRevD.105.075005}{\emph{Phys. Rev. D}
  {\bfseries 105} (2022) 075005}
  [\href{https://arxiv.org/abs/2112.15214}{{\ttfamily 2112.15214}}].

\bibitem{Haque:2022kez}
M.R.~Haque and D.~Maity, \emph{{Gravitational reheating}},
  \href{https://doi.org/10.1103/PhysRevD.107.043531}{\emph{Phys. Rev. D}
  {\bfseries 107} (2023) 043531}
  [\href{https://arxiv.org/abs/2201.02348}{{\ttfamily 2201.02348}}].

\bibitem{Barman:2023opy}
B.~Barman, N.~Bernal and J.~Rubio, \emph{{Rescuing Gravitational-Reheating in
  Chaotic Inflation}},  \href{https://arxiv.org/abs/2310.06039}{{\ttfamily
  2310.06039}}.

\bibitem{Haque:2023zhb}
M.R.~Haque, D.~Maity and R.~Mondal, \emph{{$\nu$GRe: Gravitational $\nu$trino
  Reheating}},  \href{https://arxiv.org/abs/2311.07684}{{\ttfamily
  2311.07684}}.

\bibitem{Dufaux:2006ee}
J.F.~Dufaux, G.N.~Felder, L.~Kofman, M.~Peloso and D.~Podolsky,
  \emph{{Preheating with trilinear interactions: Tachyonic resonance}},
  \href{https://doi.org/10.1088/1475-7516/2006/07/006}{\emph{JCAP} {\bfseries
  07} (2006) 006} [\href{https://arxiv.org/abs/hep-ph/0602144}{{\ttfamily
  hep-ph/0602144}}].

\bibitem{Maity:2018qhi}
D.~Maity and P.~Saha, \emph{{(P)reheating after minimal Plateau Inflation and
  constraints from CMB}},
  \href{https://doi.org/10.1088/1475-7516/2019/07/018}{\emph{JCAP} {\bfseries
  07} (2019) 018} [\href{https://arxiv.org/abs/1811.11173}{{\ttfamily
  1811.11173}}].

\bibitem{Co:2022bgh}
R.T.~Co, Y.~Mambrini and K.A.~Olive, \emph{{Inflationary gravitational
  leptogenesis}},
  \href{https://doi.org/10.1103/PhysRevD.106.075006}{\emph{Phys. Rev. D}
  {\bfseries 106} (2022) 075006}
  [\href{https://arxiv.org/abs/2205.01689}{{\ttfamily 2205.01689}}].

\bibitem{Barman:2022qgt}
B.~Barman, S.~Cl\'ery, R.T.~Co, Y.~Mambrini and K.A.~Olive, \emph{{Gravity as a
  portal to reheating, leptogenesis and dark matter}},
  \href{https://doi.org/10.1007/JHEP12(2022)072}{\emph{JHEP} {\bfseries 12}
  (2022) 072} [\href{https://arxiv.org/abs/2210.05716}{{\ttfamily
  2210.05716}}].

\bibitem{Clery:2022wib}
S.~Clery, Y.~Mambrini, K.A.~Olive, A.~Shkerin and S.~Verner,
  \emph{{Gravitational portals with nonminimal couplings}},
  \href{https://doi.org/10.1103/PhysRevD.105.095042}{\emph{Phys. Rev. D}
  {\bfseries 105} (2022) 095042}
  [\href{https://arxiv.org/abs/2203.02004}{{\ttfamily 2203.02004}}].

\bibitem{Xu:2023lxw}
Y.~Xu, \emph{{Constraining axion and ALP dark matter from misalignment during
  reheating}}, \href{https://doi.org/10.1103/PhysRevD.108.083536}{\emph{Phys.
  Rev. D} {\bfseries 108} (2023) 083536}
  [\href{https://arxiv.org/abs/2308.15322}{{\ttfamily 2308.15322}}].

\bibitem{Giudice:2000ex}
G.F.~Giudice, E.W.~Kolb and A.~Riotto, \emph{{Largest temperature of the
  radiation era and its cosmological implications}},
  \href{https://doi.org/10.1103/PhysRevD.64.023508}{\emph{Phys. Rev. D}
  {\bfseries 64} (2001) 023508}
  [\href{https://arxiv.org/abs/hep-ph/0005123}{{\ttfamily hep-ph/0005123}}].

\bibitem{Planck:2018vyg}
{\scshape Planck} collaboration, \emph{{Planck 2018 results. VI. Cosmological
  parameters}},
  \href{https://doi.org/10.1051/0004-6361/201833910}{\emph{Astron. Astrophys.}
  {\bfseries 641} (2020) A6}
  [\href{https://arxiv.org/abs/1807.06209}{{\ttfamily 1807.06209}}].

\bibitem{Mambrini:2022uol}
Y.~Mambrini, K.A.~Olive and J.~Zheng, \emph{{Post-inflationary dark matter
  bremsstrahlung}},
  \href{https://doi.org/10.1088/1475-7516/2022/10/055}{\emph{JCAP} {\bfseries
  10} (2022) 055} [\href{https://arxiv.org/abs/2208.05859}{{\ttfamily
  2208.05859}}].

\bibitem{Hooper:2018buz}
D.~Hooper, G.~Krnjaic, A.J.~Long and S.D.~Mcdermott, \emph{{Can the Inflaton
  Also Be a Weakly Interacting Massive Particle?}},
  \href{https://doi.org/10.1103/PhysRevLett.122.091802}{\emph{Phys. Rev. Lett.}
  {\bfseries 122} (2019) 091802}
  [\href{https://arxiv.org/abs/1807.03308}{{\ttfamily 1807.03308}}].

\bibitem{Garcia:2021gsy}
M.A.G.~Garcia, Y.~Mambrini, K.A.~Olive and S.~Verner, \emph{{On the Realization
  of WIMPflation}},
  \href{https://doi.org/10.1088/1475-7516/2021/10/061}{\emph{JCAP} {\bfseries
  10} (2021) 061} [\href{https://arxiv.org/abs/2107.07472}{{\ttfamily
  2107.07472}}].

\bibitem{BICEP:2021xfz}
{\scshape BICEP, Keck} collaboration, \emph{{Improved Constraints on Primordial
  Gravitational Waves using Planck, WMAP, and BICEP/Keck Observations through
  the 2018 Observing Season}},
  \href{https://doi.org/10.1103/PhysRevLett.127.151301}{\emph{Phys. Rev. Lett.}
  {\bfseries 127} (2021) 151301}
  [\href{https://arxiv.org/abs/2110.00483}{{\ttfamily 2110.00483}}].

\bibitem{Aggarwal:2020olq}
N.~Aggarwal et~al., \emph{{Challenges and opportunities of gravitational-wave
  searches at MHz to GHz frequencies}},
  \href{https://doi.org/10.1007/s41114-021-00032-5}{\emph{Living Rev. Rel.}
  {\bfseries 24} (2021) 4} [\href{https://arxiv.org/abs/2011.12414}{{\ttfamily
  2011.12414}}].

\bibitem{LISA:2017pwj}
{\scshape LISA} collaboration, \emph{{Laser Interferometer Space Antenna}},
  \href{https://arxiv.org/abs/1702.00786}{{\ttfamily 1702.00786}}.

\bibitem{Punturo:2010zz}
M.~Punturo et~al., \emph{{The Einstein Telescope: A third-generation
  gravitational wave observatory}},
  \href{https://doi.org/10.1088/0264-9381/27/19/194002}{\emph{Class. Quant.
  Grav.} {\bfseries 27} (2010) 194002}.

\bibitem{Hild:2010id}
S.~Hild et~al., \emph{{Sensitivity Studies for Third-Generation Gravitational
  Wave Observatories}},
  \href{https://doi.org/10.1088/0264-9381/28/9/094013}{\emph{Class. Quant.
  Grav.} {\bfseries 28} (2011) 094013}
  [\href{https://arxiv.org/abs/1012.0908}{{\ttfamily 1012.0908}}].

\bibitem{Sathyaprakash:2012jk}
B.~Sathyaprakash et~al., \emph{{Scientific Objectives of Einstein Telescope}},
  \href{https://doi.org/10.1088/0264-9381/29/12/124013}{\emph{Class. Quant.
  Grav.} {\bfseries 29} (2012) 124013}
  [\href{https://arxiv.org/abs/1206.0331}{{\ttfamily 1206.0331}}].

\bibitem{Maggiore:2019uih}
M.~Maggiore et~al., \emph{{Science Case for the Einstein Telescope}},
  \href{https://doi.org/10.1088/1475-7516/2020/03/050}{\emph{JCAP} {\bfseries
  03} (2020) 050} [\href{https://arxiv.org/abs/1912.02622}{{\ttfamily
  1912.02622}}].

\bibitem{Crowder:2005nr}
J.~Crowder and N.J.~Cornish, \emph{{Beyond LISA: Exploring future gravitational
  wave missions}},
  \href{https://doi.org/10.1103/PhysRevD.72.083005}{\emph{Phys. Rev. D}
  {\bfseries 72} (2005) 083005}
  [\href{https://arxiv.org/abs/gr-qc/0506015}{{\ttfamily gr-qc/0506015}}].

\bibitem{Corbin:2005ny}
V.~Corbin and N.J.~Cornish, \emph{{Detecting the cosmic gravitational wave
  background with the big bang observer}},
  \href{https://doi.org/10.1088/0264-9381/23/7/014}{\emph{Class. Quant. Grav.}
  {\bfseries 23} (2006) 2435}
  [\href{https://arxiv.org/abs/gr-qc/0512039}{{\ttfamily gr-qc/0512039}}].

\bibitem{Harry:2006fi}
G.M.~Harry, P.~Fritschel, D.A.~Shaddock, W.~Folkner and E.S.~Phinney,
  \emph{{Laser interferometry for the big bang observer}},
  \href{https://doi.org/10.1088/0264-9381/23/15/008}{\emph{Class. Quant. Grav.}
  {\bfseries 23} (2006) 4887}.

\bibitem{Seto:2001qf}
N.~Seto, S.~Kawamura and T.~Nakamura, \emph{{Possibility of direct measurement
  of the acceleration of the universe using 0.1-Hz band laser interferometer
  gravitational wave antenna in space}},
  \href{https://doi.org/10.1103/PhysRevLett.87.221103}{\emph{Phys. Rev. Lett.}
  {\bfseries 87} (2001) 221103}
  [\href{https://arxiv.org/abs/astro-ph/0108011}{{\ttfamily
  astro-ph/0108011}}].

\bibitem{Kudoh:2005as}
H.~Kudoh, A.~Taruya, T.~Hiramatsu and Y.~Himemoto, \emph{{Detecting a
  gravitational-wave background with next-generation space interferometers}},
  \href{https://doi.org/10.1103/PhysRevD.73.064006}{\emph{Phys. Rev. D}
  {\bfseries 73} (2006) 064006}
  [\href{https://arxiv.org/abs/gr-qc/0511145}{{\ttfamily gr-qc/0511145}}].

\bibitem{Li:2003tv}
F.-Y.~Li, M.-X.~Tang and D.-P.~Shi, \emph{{Electromagnetic response of a
  Gaussian beam to high frequency relic gravitational waves in quintessential
  inflationary models}},
  \href{https://doi.org/10.1103/PhysRevD.67.104008}{\emph{Phys. Rev. D}
  {\bfseries 67} (2003) 104008}
  [\href{https://arxiv.org/abs/gr-qc/0306092}{{\ttfamily gr-qc/0306092}}].

\bibitem{Berlin:2021txa}
A.~Berlin, D.~Blas, R.~Tito~D'Agnolo, S.A.R.~Ellis, R.~Harnik, Y.~Kahn et~al.,
  \emph{{Detecting high-frequency gravitational waves with microwave
  cavities}}, \href{https://doi.org/10.1103/PhysRevD.105.116011}{\emph{Phys.
  Rev. D} {\bfseries 105} (2022) 116011}
  [\href{https://arxiv.org/abs/2112.11465}{{\ttfamily 2112.11465}}].

\bibitem{Berlin:2023grv}
A.~Berlin, D.~Blas, R.~Tito~D'Agnolo, S.A.R.~Ellis, R.~Harnik, Y.~Kahn et~al.,
  \emph{{Electromagnetic cavities as mechanical bars for gravitational waves}},
  \href{https://doi.org/10.1103/PhysRevD.108.084058}{\emph{Phys. Rev. D}
  {\bfseries 108} (2023) 084058}
  [\href{https://arxiv.org/abs/2303.01518}{{\ttfamily 2303.01518}}].

\bibitem{Herman:2022fau}
N.~Herman, L.~Lehoucq and A.~F\'{u}zfa, \emph{{Electromagnetic antennas for the
  resonant detection of the stochastic gravitational wave background}},
  \href{https://doi.org/10.1103/PhysRevD.108.124009}{\emph{Phys. Rev. D}
  {\bfseries 108} (2023) 124009}
  [\href{https://arxiv.org/abs/2203.15668}{{\ttfamily 2203.15668}}].

\bibitem{Armengaud:2014gea}
E.~Armengaud et~al., \emph{{Conceptual Design of the International Axion
  Observatory (IAXO)}},
  \href{https://doi.org/10.1088/1748-0221/9/05/T05002}{\emph{JINST} {\bfseries
  9} (2014) T05002} [\href{https://arxiv.org/abs/1401.3233}{{\ttfamily
  1401.3233}}].

\bibitem{IAXO:2019mpb}
{\scshape IAXO} collaboration, \emph{{Physics potential of the International
  Axion Observatory (IAXO)}},
  \href{https://doi.org/10.1088/1475-7516/2019/06/047}{\emph{JCAP} {\bfseries
  06} (2019) 047} [\href{https://arxiv.org/abs/1904.09155}{{\ttfamily
  1904.09155}}].

\bibitem{Caprini:2018mtu}
C.~Caprini and D.G.~Figueroa, \emph{{Cosmological Backgrounds of Gravitational
  Waves}}, \href{https://doi.org/10.1088/1361-6382/aac608}{\emph{Class. Quant.
  Grav.} {\bfseries 35} (2018) 163001}
  [\href{https://arxiv.org/abs/1801.04268}{{\ttfamily 1801.04268}}].

\bibitem{COrE:2011bfs}
{\scshape COrE} collaboration, \emph{{COrE (Cosmic Origins Explorer) A White
  Paper}},  \href{https://arxiv.org/abs/1102.2181}{{\ttfamily 1102.2181}}.

\bibitem{EUCLID:2011zbd}
{\scshape EUCLID} collaboration, \emph{{Euclid Definition Study Report}},
  \href{https://arxiv.org/abs/1110.3193}{{\ttfamily 1110.3193}}.

\bibitem{Ben-Dayan:2019gll}
I.~Ben-Dayan, B.~Keating, D.~Leon and I.~Wolfson, \emph{{Constraints on scalar
  and tensor spectra from $N_{eff}$}},
  \href{https://doi.org/10.1088/1475-7516/2019/06/007}{\emph{JCAP} {\bfseries
  06} (2019) 007} [\href{https://arxiv.org/abs/1903.11843}{{\ttfamily
  1903.11843}}].

\bibitem{Maggiore:1999vm}
M.~Maggiore, \emph{{Gravitational wave experiments and early universe
  cosmology}}, \href{https://doi.org/10.1016/S0370-1573(99)00102-7}{\emph{Phys.
  Rept.} {\bfseries 331} (2000) 283}
  [\href{https://arxiv.org/abs/gr-qc/9909001}{{\ttfamily gr-qc/9909001}}].

\end{thebibliography}\endgroup
	%%%%%%%%%%%%%%%%%%%%%%%%%%%%%%%%%%%%%%%%%%
\end{document}